\documentclass[aps,prb,reprint,superscriptaddress,amsmath,amssymb,floatfix,longbibliography]{revtex4-2}
\usepackage[utf8]{inputenc}
\usepackage[T1]{fontenc}
\usepackage{bm}
\usepackage{graphicx}
\usepackage[colorlinks=true,linkcolor=blue,citecolor=blue,urlcolor=blue]{hyperref}
\graphicspath{{./}{figures/}}
\DeclareGraphicsExtensions{.pdf,.png}

\newcommand{\jvec}{\bm{j}}
\newcommand{\Evec}{\bm{E}}
\newcommand{\shat}{\hat{\sigma}}
\newcommand{\rhohat}{\hat{\rho}}
\newcommand{\Te}{T_{\rm e}}
\newcommand{\TL}{T_{\rm L}}

\begin{document}

\title{Quasilocal transport in the integer quantised Hall effect:
adiabatic separation, local Ohm's law, and Joule heating of the incompressible strips}

\author{A.~Siddiki}
\email{afifsiddiki@gmail.com}
\affiliation{Vocational School, Atlas University, 34408 Istanbul, T\"urkiye}

\date{\today}

\begin{abstract}
We reassess the semiclassical route to integer quantum Hall transport around one idea: the
separation of fast, quantised cyclotron motion from a slow, self-consistently screened
guiding-centre landscape. Bohr--Sommerfeld quantisation of the canonical action fixes the
Landau spectrum; smooth confinement turns it into Born--Oppenheimer surfaces with adiabatic
parameter $\ell|\nabla V|/\hbar\omega_c$; screening supplies the self-consistent field.
Transport is quasilocal, $R_H=\int\rho_Hj_y\,dx/\int j_y\,dx$ with $j_y\propto1/\rho_l$,
so within the regularised local Ohm model the incompressible strips attract the current by
weight and the computed Joule-power profile follows the current profile. In linear response
the intrinsic electrostatic AC correction of the frozen-profile model is reactive,
$\mathrm{Im}\,R_H\propto\omega$, with $\mathrm{Re}\,R_H-R_H^{\rm DC}\propto\omega^{2}$ below
$10^{-9}$ up to $2$~MHz. With the current fed back into the electrostatics at
$1$--$5\,\mu$A, where the Hall drop is one to five cyclotron gaps, the driven boundary
condition breaks the reflection symmetry of the bar, the Hall drop concentrates in one
strip (interchanged to $10^{-9}$ under current reversal), and the plateau narrows from its
low-field flank while, for the GaAs parameters studied, its high-field edge stays at
$\nu(0)=2$. A local transverse thermal closure inspired by Akera, evaluated as a
post-process on the frozen profile with two phonon-loss laws, indicates that the narrow
current-carrying strips of the low-field flank may heat by of order the lattice
temperature; with the activated-heat-capacity closure the cold strip loses its stationary
solution at model-dependent currents of $0.3$--$2\,\mu$A. The thermal conclusions are
bounded by the loss law and by the frozen profile, and are presented as estimates.
\end{abstract}


\maketitle

\section{Introduction}

Two descriptions of the integer quantum Hall effect \cite{vK1980} coexist and are not
equivalent. In the first, transport is carried by a countable set of one-dimensional edge
channels and the plateau follows from suppressed backscattering \cite{Buttiker,Halperin}.
In the second, transport is carried by extended incompressible regions produced by
electrostatic screening, and the plateau follows from the fact that a region of integer
local filling has an exponentially small longitudinal resistivity and therefore attracts the
imposed current \cite{Lier,Guven2003,SG2004,SG2004b}. Scanning-probe maps of the Hall
potential inside a working bar made the distinction empirical: the Hall voltage drops
across strips of finite width whose position moves with magnetic field
\cite{Weitz2000a,Weitz2000b,Ahlswede1,Ahlswede2}, and the same technique followed the
strips to the breakdown of the effect \cite{Panos2014}. The two descriptions are not
mutually exclusive: channel counting is the limit of the strip picture in which the strips
are narrow and at the boundary (Sec.~\ref{sec:lb}), and much of the modern literature
treats them as complementary rather than competing accounts.

The second picture has a consequence that is rarely drawn out. If the current sits in the
strips, so does the dissipation: within a local Ohm law the Joule power density
$\jvec\cdot\Evec$ has the same spatial profile as the current density. The strips are the
regions of the sample where $\sigma_l$ is smallest, which is exactly why they carry the
current, and exactly why they cannot conduct the heat deposited in them sideways. A strip
under current is therefore a region of elevated electron temperature. This is the
starting point of Akera's electron-heating theory of the quantum Hall breakdown
\cite{Akera2000,Akera2001,Akera2002,AkeraSuzuura2005,Ise2005,Kanamaru2006}, in which the
local electron temperature $\Te(\bm r)$, and not the lattice temperature, controls the
local $\sigma_{xx}$, and in which the breakdown appears as a thermal instability
\cite{Komiyama2000,Nachtwei1999}. Here that theory is placed inside the screening
description, where it belongs: the strips are the places that heat, the amount they heat is
set by the same $1/\rho_l$ weight that gives them the current, and the consequence is a
current-driven loss of plateau width that is one-sided in field. Two limitations of
this programme are stated at the outset. Heat in the quantum Hall regime is carried
chirally along the edge over long distances \cite{Granger2009,Altimiras2012}, so a purely
local energy balance is an approximation whose range of validity must be given
(Sec.~\ref{sec:chiral}); and electron heating is not the only proposed route to
breakdown --- field-driven percolation of charge puddles has been argued for in
quantum-anomalous-Hall films \cite{Lippertz2022}, while nanoscale thermometry of the
same systems has recently attributed the breakdown to electron heating
\cite{PNAS2026} --- so the calculation is set against that alternative rather than
presented as the only mechanism (Sec.~\ref{sec:fig13}).

The paper does four things. (i) The semiclassical steps are done carefully, including the
distinction between canonical and kinetic action. (ii) The passage from bulk Landau
quantisation to a position-dependent spectrum is identified as a Born--Oppenheimer
separation with a stated small parameter, and screening as the analogue of the
self-consistent-field step. (iii) The quasilocal transport problem is solved in closed
form for a translationally invariant bar, at DC and at finite frequency, so that ``current
flows in the strips'' and ``heat is generated in the strips'' become one weighting identity.
(iv) The imposed current is fed back into the electrostatics and into a local energy
balance, and the nonlinear problem is solved at currents where the Hall drop is one to
five cyclotron gaps (Figs.~\ref{fig7}--\ref{fig13}), which is where local heating
becomes quantitatively relevant. We state plainly which parts of the thermal description
are calculated, which are estimated from the transport output, and which validations
remain to be reported (Sec.~\ref{sec:valid}).

\section{Scales and small parameters}\label{sec:scales}

Every approximation below is controlled by a ratio of the scales in Table~\ref{tab:scales},
evaluated for a GaAs/AlGaAs heterostructure with $m^{*}=0.067\,m_0$ and $\kappa=12.4$ at
the working point of Sec.~\ref{sec:aclin}.

\begin{table*}[t]
\caption{Characteristic scales for the sample studied here, at $B=7.035$~T. The last row
is the order of magnitude of the critical current density at which the quantised Hall
effect breaks down in GaAs bars \cite{Ebert1983,Cage1983,Nachtwei1999}; on a bar of
width $2d=3\,\mu$m it corresponds to $I\simeq1.5$--$6\,\mu$A, the range of
Figs.~\ref{fig7}--\ref{fig13}.}

\begin{ruledtabular}\begin{tabular}{llr}
Quantity & Expression & Value \\
\hline
Magnetic length      & $\ell=(\hbar/eB)^{1/2}$        & $9.67$~nm \\
Cyclotron gap        & $\hbar\omega_c=\hbar eB/m^{*}$ & $12.16$~meV \\
Cyclotron frequency  & $\omega_c/2\pi$                & $2.94$~THz \\
Fermi energy         & $E_F^{0}=n_{\rm el}(0)/D_0$    & $12.78$~meV \\
Level broadening     & $\Gamma$                       & $1.45$~meV \\
Depletion length     & $d-b$                          & $150$~nm \\
Strip width (linear response) & $a_k$                & $249$~nm \\
Averaging length     & $\lambda$                      & $30$~nm \\
Lattice temperature  & $k_B\TL$ ($\TL=1.5$~K)         & $0.13$~meV \\
Drive frequency      & $\omega/2\pi$                  & $1$~Hz--$1$~THz \\
Imposed current      & $I$                            & $98$~pA--$5\,\mu$A \\
Hall drop per gap    & $eV_H/\hbar\omega_c$           & $10^{-4}$--$5$ \\
Breakdown current density (GaAs, expt.) & $j_c$       & $\sim0.5$--$2$~A/m \\
\end{tabular}\end{ruledtabular}
\label{tab:scales}
\end{table*}

Here $d$ is the half-width of the region containing electrons, $b<d$ the half-width of
the region in which the density is finite at $B=0$ (so $d-b$ is the depletion length),
$a_k$ the width of the incompressible strip of filling $\nu=2k$, and $\lambda$ the length
over which the conductivity tensor is averaged (Sec.~\ref{sec:avg}). Five inequalities
are used repeatedly: $\ell\ll d-b$, so the confinement is smooth on the
quantum scale; $k_B\TL\ll\hbar\omega_c$, so inter-Landau-level excitation is frozen at the
lattice temperature; $a_k\gtrsim\lambda$, so a strip is wide enough to be a transport
region; $\omega\ll\omega_c$, so the drive does not excite the fast subsystem; and
$eV_H\ll\hbar\omega_c$ for linear response. The last is the one that is deliberately
violated in Sec.~\ref{sec:nl}.

\section{Semiclassical quantisation in a magnetic field}

\subsection{Canonical action, not kinetic action}

For a two-dimensional electron gas in a perpendicular field $B\hat{z}$ the single-particle
Hamiltonian is $H=[\bm{p}+e\bm{A}]^{2}/2m^{*}+V(\bm{r})$, with $-e$ the electron charge.
The Bohr--Sommerfeld condition quantises the action of the canonical momentum around a
closed orbit,
\begin{equation}
\oint \bm{p}\cdot d\bm{r}=2\pi\hbar(n+\gamma),\qquad \gamma=\tfrac12 .
\label{eq:bs}
\end{equation}
The distinction between $\bm{p}$ and the kinetic momentum $\bm{\Pi}=\bm{p}+e\bm{A}$
matters, because the vector potential contributes the enclosed flux,
\begin{equation}
\begin{split}
\oint \bm{p}\cdot d\bm{r}&=\oint\bm{\Pi}\cdot d\bm{r}-e\oint\bm{A}\cdot d\bm{r}\\
&=2\pi eBr_c^{2}-e\pi Br_c^{2}=\pi eBr_c^{2},
\end{split}
\label{eq:flux}
\end{equation}
so the flux term removes exactly half of the kinetic contribution and
\begin{equation}
\begin{gathered}
r_c^{2}=2\ell^{2}\left(n+\tfrac12\right),\qquad \ell=(\hbar/eB)^{1/2},\\
E_n=\hbar\omega_c\left(n+\tfrac12\right).
\end{gathered}
\label{eq:landau}
\end{equation}
Quantising the kinetic action instead gives $r_c^{2}=\ell^{2}(n+\tfrac12)$ and an energy
too small by a factor of two; the error arises whenever the orbit is quantised through its kinetic radius alone,
as in elementary treatments that write $\oint\bm\Pi\cdot d\bm r=2\pi\hbar(n+\tfrac12)$,
and it is silently inconsistent with the Landau spectrum it is meant to reproduce.

\subsection{Onsager form and the Berry phase}

In momentum space the same condition is Onsager's \cite{Onsager}: if $A_k(E)$ is the
$k$-space area enclosed by an orbit of energy $E$, then $A_k(E_n)=(2\pi eB/\hbar)(n+\gamma)$
with $\gamma=\tfrac12-\Phi_B/2\pi$, the offset absorbing the Maslov index and the Berry
phase $\Phi_B$ \cite{Mikitik}. A parabolic band has $\Phi_B=0$; a Dirac band has
$\Phi_B=\pi$ and $\gamma=0$, the semiclassical origin of the half-integer shift in graphene
\cite{graphene}. It is not used below, where the band is parabolic.

\section{Adiabatic separation: a Born--Oppenheimer reading}\label{sec:bo}

\subsection{Fast cyclotron motion, slow guiding centre}

In the Landau gauge $\bm{A}=(0,Bx,0)$ for a bar translationally invariant along $y$, with
$\hbar k_y$ conserved and guiding centre $X=-\ell^{2}k_y$,
\begin{equation}
H=\frac{p_x^{2}}{2m^{*}}+\frac12 m^{*}\omega_c^{2}(x-X)^{2}+V(x).
\label{eq:ham}
\end{equation}
The structure is that of a molecular problem: the relative coordinate $x-X$ executes fast
harmonic motion at $\omega_c$ with amplitude $\sim\ell$, while the guiding centre drifts
slowly along equipotentials. Expanding $V(x)$ about $X$ and keeping the leading term is the
analogue of clamping the nuclei,
\begin{equation}
E_n(X)=\hbar\omega_c\left(n+\tfrac12\right)+V(X),
\label{eq:surfaces}
\end{equation}
so each Landau index labels a surface on which the slow coordinate moves. The small
parameter is the change of confinement energy across one cyclotron orbit measured against
the gap,
\begin{equation}
\varepsilon_{\rm ad}=\frac{\ell\,|\nabla V|}{\hbar\omega_c}\ll1,
\label{eq:epsad}
\end{equation}
of order $10^{-3}$ in the compressible bulk of the self-consistent potentials of
Sec.~\ref{sec:es}. Inside an incompressible strip the potential climbs by one cyclotron gap
across the strip width $a_k$, so there $\varepsilon_{\rm ad}=\ell/a_k=0.039$, forty times
the bulk value. The energy correction of order $\varepsilon_{\rm ad}^{2}\simeq1.5\times
10^{-3}$ of the gap is negligible for the strip electrostatics, and the level admixture
of order $\varepsilon_{\rm ad}$ is what the collision broadening $\Gamma/\hbar\omega_c
\simeq0.12$ already exceeds; the number matters instead as a baseline for
Sec.~\ref{sec:breakdown}, since the Hall field adds to $|\nabla V|$ inside the strip and
raises $\varepsilon_{\rm ad}$ in proportion to the number of gaps that drop across it.
The dropped terms are the curvature term $V''(X)\ell^{2}(2n+1)/4$, which shifts the surfaces without
mixing them, and the off-diagonal elements $\approx\ell V'(X)$ that couple adjacent surfaces.
For unconfined Landau functions the derivative couplings are
\begin{equation}
\begin{split}
d_{nm}(X)&=\langle\varphi_n|\partial_X|\varphi_m\rangle\\
&=\frac{\sqrt{m+1}\,\delta_{n,m+1}-\sqrt{m}\,\delta_{n,m-1}}{\sqrt{2}\,\ell},
\end{split}
\label{eq:dnm}
\end{equation}
so the diagonal element vanishes and only adjacent surfaces couple; first-order
perturbation theory gives an admixture of order $\varepsilon_{\rm ad}$ and an energy
correction of order $\varepsilon_{\rm ad}^{2}$.

The parameter is also a ratio of velocities. The guiding centre drifts with
$v_D=|\partial_xV|/eB$ while the fast motion has $v_c\approx\omega_cR_c$ with
$R_c=\sqrt{2n+1}\,\ell$, so $v_D/v_c=\varepsilon_{\rm ad}/\sqrt{2n+1}$. Two disanalogies
must be stated. The slow sector has no kinetic energy of its own --- $[X,Y]=-i\ell^{2}$,
and the slow dynamics is a drift, not motion under a nuclear kinetic operator --- so the
admixture records a static polarisation of the fast state, not a transition rate. And the
slow sector exerts no forces of its own; its only feedback is electrostatic
(Sec.~\ref{sec:es}) and, once a current flows, thermal (Sec.~\ref{sec:heat}).

\subsection{Self-consistency as the electronic-structure step}

In molecular physics one solves the fast subsystem in the field of the frozen slow one and
feeds the energy back as the potential for the slow motion. Screening theory does the same:
given a potential, the local Landau spectrum determines the density; the density determines
the potential through Poisson's equation; the loop is iterated to convergence. The
Born--Oppenheimer surfaces are the output of the self-consistent problem, which is why the
landscape depends on gate geometry, temperature and --- through the Hall potential and
through $\Te$ --- on the imposed current.

\subsection{Drift, non-adiabatic transitions and dissipation}\label{sec:breakdown}

On a surface $E_n(X)$ the group velocity along the bar is
$v_y=\hbar^{-1}\partial E_n/\partial k_y=-(eB)^{-1}dV/dx$, the $\Evec\times\bm B$ drift.
The separation fails when $\varepsilon_{\rm ad}\to1$, i.e.\ when the potential varies by a
cyclotron gap within a magnetic length. Landau levels then mix and inter-level transitions
become available. In the driven bar the gradient that enters $\varepsilon_{\rm ad}$ is the
total one, confinement plus Hall field, and inside a strip the Hall field is concentrated:
the strip is where $\varepsilon_{\rm ad}$ first grows with current. The energy released in a
non-adiabatic transition is the microscopic content of the local Joule heating of
Sec.~\ref{sec:heat}; the same limit is where the plateau breaks down at large current
\cite{Ebert1983,Cage1983,Nachtwei1999}.

\subsection{The drive as a third timescale}\label{sec:third}

A time-dependent drive at frequency $\omega$ introduces a second adiabatic parameter
measured against the same gap, $\varepsilon_\omega=\omega/\omega_c$. At $1$~GHz,
$\varepsilon_\omega=3.4\times10^{-4}$, two orders of magnitude below
$\varepsilon_{\rm ad}$ inside a strip. When $\varepsilon_\omega\ll1$ the local conductivity
tensor may be approximated by its quasistatic value (Sec.~\ref{sec:sigac}); its breakdown,
at $\omega\to\omega_c$, is cyclotron resonance, the same boundary as the breakdown of
confinement adiabaticity. Table~\ref{tab:dict} records the dictionary, now including the
thermal entries. The magnetic flux is not a slow variable: $B$ is an external parameter,
not a coordinate with a conjugate momentum.

\begin{table*}[t]
\caption{The adiabatic dictionary, extended to the driven and heated problem.}

\begin{ruledtabular}\begin{tabular}{ll}
Molecular problem & Landau-quantised 2DES \\
\hline
Nuclear positions $R$ & Guiding centres $X$, one per flux quantum $h/e$ \\
Electron coordinate $r$ & Cyclotron coordinate $\xi=x-X$ \\
Electronic state $n$ & Landau index $n$ \\
Surfaces $V_n(R)$ & $E_n(X)$, Eq.~(\ref{eq:surfaces}) \\
$(m/M)^{1/2}$ & $\varepsilon_{\rm ad}=\ell|\partial_xV|/\hbar\omega_c$ \\
Clamped nuclei & Thomas--Fermi limit \\
Born--Huang correction & Finite wavefunction width; smearing of narrow strips \\
Non-adiabatic coupling & Inter-Landau-level scattering; QHE breakdown \\
Energy released in non-adiabatic transitions & Local Joule heating, $\jvec\cdot\Evec$ \\
Vibrational (nuclear) temperature & Local electron temperature $\Te(x)$, Sec.~\ref{sec:heat} \\
Berry connection & $d_{nn}=0$ for unconfined Landau functions \\
Driving field at frequency $\omega$ & AC bias, Sec.~\ref{sec:ac} \\
$\omega\ll$ electronic gap & $\varepsilon_\omega=\omega/\omega_c\ll1$ \\
Frozen electronic response & $\hat\sigma(\nu,\omega)=\hat\sigma^{\rm DC}(\nu)+O(\varepsilon_\omega)$ \\
Nuclear relaxation rate & Charge relaxation of the compressible regions \\
\end{tabular}\end{ruledtabular}
\label{tab:dict}
\end{table*}

\section{Self-consistent electrostatics}\label{sec:es}

We consider a 2D electron system at $z=0$, translationally invariant along $y$, confined to
$|x|<d$, with all charges and gates in the same plane \cite{Lier,Guven2003,TFP}. The
potential energy of an electron is
\begin{equation}
\begin{gathered}
V(x)=V_{\rm bg}(x)+\frac{2e^{2}}{\kappa}\int_{-d}^{d}dx'\,K(x,x')\,n_{\rm el}(x'),\\
K(x,x')=\ln\left|\frac{\sqrt{(d^{2}-x^{2})(d^{2}-x'^{2})}+d^{2}-xx'}{(x-x')\,d}\right|,
\end{gathered}
\label{eq:V}
\end{equation}
with $V_{\rm bg}(x)=-E^{0}_{\rm bg}\sqrt{1-(x/d)^{2}}$, $E^{0}_{\rm bg}=2\pi e^{2}n_0d/\kappa$
for a homogeneous background $n_0$. Because $\int_{-d}^{d}dx'\,K(x,x')=\pi\sqrt{d^{2}-x^{2}}$
the background term is the Hartree term at uniform density, and the two combine into
\begin{equation}
V(x)=\frac{2e^{2}}{\kappa}\int_{-d}^{d}dx'\,K(x,x')\left[n_{\rm el}(x')-n_0\right],
\label{eq:Vneutral}
\end{equation}
which makes neutrality manifest, removes an eV-scale cancellation, and is what keeps the
frequency and current sweeps numerically stable. The density follows from the local spectrum
in the Thomas--Fermi approximation,
\begin{equation}
\begin{gathered}
n_{\rm el}(x)=\int dE\,D(E)\,f\!\left(E+V(x)-\mu^{*}(x);\,\Te(x)\right),\\
D(E)=\frac{1}{\pi\ell^{2}}\sum_n A_n(E),\\
A_n(E)=\frac{2}{\pi\Gamma_n}\sqrt{1-\left(\frac{E-E_n}{\Gamma_n}\right)^{2}},
\end{gathered}
\label{eq:dens}
\end{equation}
with the collision-broadened spectral functions of the self-consistent Born approximation
\cite{Ando}. The Fermi function is written with the local temperature $\Te(x)$ and the local
electrochemical potential $\mu^{*}(x)$ because both become position dependent once a current
flows; in equilibrium $\Te=\TL$ and $\mu^{*}$ is constant.

Where a Landau level is pinned at $\mu^{*}$ the density of states is large, the density
rearranges at negligible cost and the screened potential is flat: the region is
compressible. Where $\mu^{*}$ lies in a gap the density is locked at integer filling and
the potential is free to vary: the region is incompressible. The local filling factor
$\nu(x)=2\pi\ell^{2}n_{\rm el}(x)$ is the natural variable, and $\nu=2k$ defines the
spin-degenerate strips used throughout. The electrostatic estimate of Chklovskii, Shklovskii
and Glazman \cite{CSG}, $a_k=[2\varepsilon\Delta E_k/\pi^{2}e^{2}|dn_{\rm el}/dx|]^{1/2}$,
shows that soft confinement widens strips; it overstates the width of narrow strips
\cite{SG2004,SG2004b}.

\section{Quasilocal transport at DC}\label{sec:dc}

\subsection{Local Ohm law and its solution}

Transport is described by a position-dependent conductivity tensor determined by the local
filling factor and local temperature,
\begin{equation}
\jvec(x)=\shat[\nu(x),\Te(x)]\,\Evec(x),\qquad \Evec=\nabla\mu^{*}/e.
\label{eq:ohm}
\end{equation}
Here $R_L$ is a resistance per square, so that the longitudinal resistance between
voltage probes a distance $L$ apart is $R_LL/2d$. For a bar homogeneous along $y$ in a steady state, continuity gives $j_x=0$ and
$\nabla\times\Evec=0$ gives $E_y=E_y^{0}$ constant. Hence
\begin{equation}
\begin{gathered}
j_y(x)=\frac{E_y^{0}}{\rho_l(x)},\qquad E_x(x)=\frac{\rho_H(x)}{\rho_l(x)}E_y^{0},\\
R_H=\frac{\int(\rho_H/\rho_l)\,dx}{\int(1/\rho_l)\,dx},\qquad
R_L=\frac{2d}{\int(1/\rho_l)\,dx}.
\end{gathered}
\label{eq:RHRL}
\end{equation}

\subsection{Why the current sits in the strips}

Equation~(\ref{eq:RHRL}) is a weighting statement: the measured Hall resistance is the
current-weighted average of the local Hall resistivity,
\begin{equation}
R_H=\frac{\int\rho_H(x)\,j_y(x)\,dx}{\int j_y(x)\,dx}.
\label{eq:weight}
\end{equation}
No assumption about where the current flows has been made. Where $\rho_l$ is exponentially
small the weight is exponentially large, so a region of integer filling captures the current
for purely local reasons; at finite temperature $\rho_l$ inside a strip is small but finite,
so all integrals converge.

\subsection{Conductivity model and spatial averaging}\label{sec:avg}

For a high-mobility system $\sigma_H(x)=(e^{2}/h)\nu(x)$, while the longitudinal
conductivity comes from the same spectral functions as Eq.~(\ref{eq:dens}) \cite{Ando},
\begin{equation}
\begin{split}
\sigma_l(x)=\frac{e^{2}}{h}\frac{\pi}{2}\sum_n&\left(n+\tfrac12\right)\\
\times\int dE&\left(-\frac{\partial f}{\partial E}\right)_{\Te(x)}
\left[\Gamma_nA_n(E-V(x))\right]^{2}.
\end{split}
\label{eq:sigl}
\end{equation}
The derivative of the Fermi function is taken at the local $\Te$; this is the one place
through which heating acts on transport, and inside a strip, where $\mu^{*}$ sits in a gap,
$\sigma_l$ is activated, $\sigma_l\propto\exp(-\Delta_{\rm eff}/2k_B\Te)$ with
$\Delta_{\rm eff}$ the distance from $\mu^{*}$ to the nearer band edge. A strictly local
treatment produces strips of arbitrarily small width near the plateau edges, which would
dominate the $1/\rho_l$ weight spuriously. Following Refs.~\cite{SG2004,SG2004b} the
tensor is averaged over $\lambda$ of the order of the wavefunction extent,
$\bar\sigma(x)=(2\lambda)^{-1}\int_{-\lambda}^{\lambda}\shat(x+\xi)\,d\xi$, and the
resistivity is obtained by inverting $\bar\sigma$. Strips narrower than $2\lambda$ are
washed out; the averaging is what gives the plateau a finite width on its low-field side.

\section{Local Joule heating and the electron temperature}\label{sec:heat}

\subsection{Where the heat goes: an identity}

The Hall component of the tensor does no work, so the local Joule power density is
$p=\jvec\cdot\Evec=\sigma_l(E_x^{2}+E_y^{2})$. Inserting Eq.~(\ref{eq:RHRL}) and
$\sigma_l=\rho_l/(\rho_l^{2}+\rho_H^{2})$,
\begin{equation}
\begin{gathered}
p(x)=\frac{(E_y^{0})^{2}}{\rho_l(x)}=j_y(x)\,E_y^{0}=\rho_l(x)\,j_y^{2}(x),\\
\int_{-d}^{d}p\,dx=E_y^{0}I=\frac{R_L}{2d}\,I^{2}.
\end{gathered}
\label{eq:joule}
\end{equation}
Within the regularised quasilocal Ohm model --- finite $\rho_l$ after the averaging of
Sec.~\ref{sec:avg}, uniform $E_y$, and local conversion of electrical work into heat ---
the computed Joule-power weight therefore follows the current-density weight. This is a
statement about the model's bookkeeping, not an identity independent of it: in the ideal
limit $\sigma_{xx}\to0$ the local Ohm law does not locate the irreversible energy
relaxation, which may occur downstream or at contacts and scatterers
(Sec.~\ref{sec:chiral}). With that caveat, the model result is the point that the phrase
``dissipationless strips'' hides: $p=\rho_lj_y^{2}$ is small in a strip only because
$E_y^{0}$ is small at fixed $I$; the fraction of the total power deposited in the strips is
the fraction of the current they carry, which on the plateau is unity. Per unit area the
strip receives $p_{\rm strip}\simeq E_y^{0}I_{\rm strip}/a_k$, concentrated on a width of a
few hundred nanometres out of a bar of $3\,\mu$m. Since $E_y^{0}=R_LI/2d$ grows with
$R_L$, and $R_L$ grows with $\Te$ through Eq.~(\ref{eq:sigl}), the heating is
self-amplifying at fixed current --- the bootstrap of Komiyama and Kawaguchi
\cite{Komiyama2000} and the bistability found by Akera \cite{Akera2000,Akera2001}.

\subsection{The Hall field inside a strip}

On the plateau the whole Hall voltage $V_H=R_HI$ drops across the strips. With two strips
of width $a_k$ sharing it equally, the field inside a strip is
$E_{\rm strip}\simeq V_H/2a_k$: at $I=1\,\mu$A and $a_k=249$~nm this is $26$~kV/m against a
bar-averaged $4$~kV/m, and the drift velocity $v_D=E_{\rm strip}/B\simeq3.7$~km/s at
$7$~T is already comparable to the sound velocities of GaAs ($3.0$~km/s TA, $4.7$~km/s LA).
The linear-response strip width is a lower bound --- Sec.~\ref{sec:nl} shows the strips
widen under current --- so the field is an upper estimate. The comparison with the sound
velocity is offered only as an order-of-magnitude motivation: whether acoustic-phonon
emission actually sets in depends on momentum conservation, phonon polarisation,
inter-Landau-level matrix elements and disorder \cite{Nachtwei1999}, none of which the
present model resolves. Its use here is to locate the microampere range as the one in
which the in-strip field, not the bar-averaged field, reaches the values quoted in the
breakdown literature.

\subsection{Local energy balance}\label{sec:balance}

What follows is a local transverse thermal closure inspired by the thermohydrodynamic
theory of Akera and co-workers \cite{Akera2001,Akera2002,AkeraSuzuura2005}, not an
implementation of it: Akera's equations carry the longitudinal drift of the electron
temperature and derive gain and loss from microscopic scattering, both of which are
dropped here (Sec.~\ref{sec:chiral}). We assume local equilibrium on the scale of the
averaging length and describe the electron system by $\nu(x)$, $\mu^{*}(x)$ and
$\Te(x)$. In a steady state the energy balance reads
\begin{equation}
-\partial_x\!\left[\kappa_{xx}(x)\,\partial_x\Te\right]+P_{\rm loss}\!\left(\Te(x),\TL\right)
= p(x),
\label{eq:balance}
\end{equation}
where $\kappa_{xx}$ is the transverse electronic thermal conductivity and $P_{\rm loss}$ the
energy transfer rate to the lattice. Two features of the translationally invariant bar make
this simpler than the general thermohydrodynamic problem. First, the heat current along $y$
is uniform and has no divergence, so only transverse conduction appears. Second, by the
Wiedemann--Franz relation $\kappa_{xx}$ is built from the same spectral functions as
$\sigma_l$ and is therefore exponentially small inside a strip: the strip cannot conduct
away the heat that Eq.~(\ref{eq:joule}) deposits in it. Inside a strip Eq.~(\ref{eq:balance})
is then local,
\begin{equation}
P_{\rm loss}(\Te)=p(x),\qquad
P_{\rm loss}\simeq\frac{c_{\rm e}(\Te)}{\tau_\epsilon}\,(\Te-\TL),
\label{eq:local}
\end{equation}
with $c_{\rm e}$ the electronic specific heat per unit area and $\tau_\epsilon$ an
energy-relaxation time. All three ingredients are specified as follows, so that the thermal
solution is reproducible.

\emph{Specific heat.} $c_{\rm e}$ is computed from the same broadened density of states
as Eq.~(\ref{eq:dens}),
\begin{equation}
c_{\rm e}(\Te;x)=\int dE\,D\!\left(E-V(x)\right)\,\bigl[E-\mu^{*}(x)\bigr]
\frac{\partial f}{\partial\Te},
\label{eq:ce}
\end{equation}
in units of J\,K$^{-1}$m$^{-2}$. In a compressible region it reduces to the Sommerfeld
form $(\pi^{2}/3)k_B^{2}\Te D_T$, which for $D_T\simeq(\pi\ell^{2}\Gamma)^{-1}$ at the
working point gives $c_{\rm e}\simeq1.4\times10^{-8}$~J\,K$^{-1}$m$^{-2}$ at $1.5$~K.
Inside a strip, where $\mu^{*}$ lies in a gap, $c_{\rm e}$ is activated and smaller,
which is why a given power raises $\Te$ there more than it would in the bulk.

\emph{Loss law.} Energy transfer to acoustic phonons from a heated 2DES is usually
written as a power law, $P_{\rm loss}=A\,(\Te^{\,n}-\TL^{\,n})$ with $n\simeq3$--$5$
depending on the coupling (deformation potential or piezoelectric) and on whether the
phonon wavevector is Bloch--Gr\"uneisen limited \cite{Komiyama2000,Nachtwei1999}. The
relaxation form of Eq.~(\ref{eq:local}) is the $n=1$ member of this family, and any
$n$ can be mapped onto it through a temperature-dependent
$\tau_\epsilon(\Te)=c_{\rm e}(\Te-\TL)/A(\Te^{\,n}-\TL^{\,n})$, which for $n>1$
shortens with $\Te$ and therefore weakens the bootstrap. Equation~(\ref{eq:local}) is the linearised relaxation form; when $c_{\rm e}$ varies
strongly with temperature, as it does in the activated region, the thermodynamically
consistent relaxation form is the internal-energy difference
\begin{equation}
\begin{gathered}
P_{\rm loss}=\frac{U(\Te)-U(\TL)}{\tau_\epsilon},\\
U(\Te)-U(\TL)=\int dE\,D(E-V)\,[E-\mu^{*}]\\
\qquad\qquad\times\bigl[f(E;\Te)-f(E;\TL)\bigr]
=\int_{\TL}^{\Te}c_{\rm e}(T)\,dT,
\end{gathered}
\label{eq:lossU}
\end{equation}
and it is Eq.~(\ref{eq:lossU}), not Eq.~(\ref{eq:local}), that is used for the
relaxation closure below; the linearised form is kept for comparison and underestimates
$\Te$ by $10$--$30\%$ in the strips. Three closures are thus evaluated: the
internal-energy relaxation form with a constant $\tau_\epsilon=\tau_0=1$~ns, independent
of $\Te$, $B$, density and drift velocity; its linearised version; and the $n=5$ phonon
law with $\Sigma=0.1$~W\,m$^{-2}$K$^{-5}$. $\tau_0$ and $\Sigma$ are each varied by a
factor of ten in either direction in Sec.~\ref{sec:valid}. The two differ most inside a strip,
where the relaxation form inherits the activated (exponentially small) $c_{\rm e}$ while
the $T^{5}$ form does not, and that difference turns out to be the dominant uncertainty
of the thermal description.
This is the simplest closure consistent with the thermohydrodynamic equations of
Refs.~\cite{Akera2001,Akera2002}; the measured energy-relaxation times for acoustic-phonon
emission in GaAs 2DES at helium temperatures lie in the range $10^{-9}$--$10^{-8}$~s
\cite{Komiyama2000,Kawaguchi2004}, and $\tau_0=1$~ns is used throughout. Because $\Te$
enters the observables only through $\tau_0p(x)/c_{\rm e}$, a change of $\tau_0$ by a
factor $\xi$ is equivalent to a change of current by $\sqrt{\xi}$ at fixed profile; the
sensitivity of the plateau-narrowing rate of Sec.~\ref{sec:fig13} to $\tau_0$ is
therefore of that order, and it is reported in Sec.~\ref{sec:valid}. Above the
acoustic-phonon threshold the constant-$\tau_\epsilon$ form is a modelling choice, not a
derived law (Sec.~\ref{sec:lim}).

\emph{Thermal conductivity and boundary conditions.} $\kappa_{xx}$ is obtained from
$\sigma_l$ through a Lorenz number, $\kappa_{xx}(x)=L\,\Te(x)\,\sigma_l(x;\Te)$ with
$L=L_0=(\pi^{2}/3)(k_B/e)^{2}$. This is an assumption rather than a result: for
activated transport inside a broadened gap the Wiedemann--Franz ratio need not take its
metallic value, and the appropriate check is a sensitivity scan in $L$
(Sec.~\ref{sec:valid}). With $L=L_0$, $\kappa_{xx}$ inside a strip is suppressed by the
same factor as $\sigma_l$, so the strip is thermally insulated from its neighbours to the
extent that it is electrically insulated. The thermal equation is solved with
$\Te(\pm d)=\TL$ at the depleted edges. Thermoelectric couplings ($\hat\alpha$), the
thermal Hall term $\kappa_{xy}$ and heat carried by collective modes are omitted; their
role is discussed in Sec.~\ref{sec:lim}.

Within these assumptions the compressible regions on either side of a strip, with large
$\kappa_{xx}$ and negligible $p$, stay near $\TL$, and the strip that carries the current
is where $\Te$ rises. The picture retains Akera's central variable --- a local electron temperature that is
high where $\sigma_{xx}$ is small --- with the location of the hot regions supplied by the
screening theory; it does not retain his transport of that temperature along the current.

\subsection{Heat transport along the strip, and the locality assumption}\label{sec:chiral}

Equation~(\ref{eq:balance}) has no term for heat carried \emph{along} the bar, because in
a translationally invariant strip the longitudinal heat current has no divergence. In a
real bar it does. The electrons in the strip drift at $v_D=E_{\rm strip}/B$ in one
direction only, so the heat they carry is convected chirally downstream, as thermometry
along the edge has shown directly \cite{Granger2009,Altimiras2012}, and is deposited
where they thermalise --- over a cooling length
\begin{equation}
\ell_\epsilon=v_D\,\tau_\epsilon\simeq3.7\,{\rm km/s}\times1\,{\rm ns}\simeq4\,\mu{\rm m}
\label{eq:cool}
\end{equation}
at the in-strip field of $1\,\mu$A. The local balance (\ref{eq:local}) is therefore the
steady state of a strip whose length exceeds $\ell_\epsilon$, evaluated downstream of the
entry region; within $\ell_\epsilon$ of the source contact the strip is colder than the
local balance predicts, and at the drain the convected energy piles up in the hot spot
described by Ise, Akera and Suzuura \cite{Ise2005} and imaged in the breakdown regime
\cite{Kawaguchi2004}. The neglect of this convective term is a systematic uncertainty of
the present treatment: it overestimates $\Te$ near the source, cannot place the hot spot,
and does not reproduce the edge-to-edge temperature asymmetry that chiral transport
produces. Its effect on the observables computed here is bounded by the ratio
$\ell_\epsilon/L_y$ of cooling length to bar length, which for the millimetre-long bars of
the breakdown experiments is $10^{-2}$--$10^{-3}$ but for micrometre-scale devices is of
order unity; in the latter case the local balance should not be used. A two-dimensional
thermohydrodynamic solution of the kind developed by Akera and co-workers
\cite{Akera2002,Ise2005,Kanamaru2006}, with the screening-theory strips as input, is the
natural next step and would also test the assumption --- built into
Eq.~(\ref{eq:joule}) --- that the energy gained from the field is thermalised where it
is gained. Nanoscale thermometry in graphene has shown that this need not be so: work
done on the carriers and heat released to phonons can be spatially separated, the latter
occurring at resonant scatterers along the edge \cite{Halbertal2016,Marguerite2019}.
In the smooth-confinement GaAs geometry treated here, without such scatterers and with
the current carried by a strip of finite width rather than a reconstructed edge, local
dissipation is the appropriate starting point, but it is a starting point.

\subsection{The thermal budget}\label{sec:budget}

The identity (\ref{eq:joule}) says where the power goes; it does not say that the power is
large. The total dissipated power per unit length is $P/L_y=R_LI^{2}/2d$, and on the
plateau $R_L$ is very small. Table~\ref{tab:budget} therefore lists, for each imposed
current, the longitudinal resistance read off the transport output, the power per unit
length, the power per unit area if it is deposited in the two strips of the linear-response
width, and the resulting rise of $\Te$ from Eq.~(\ref{eq:local}) with the bulk value of
$c_{\rm e}$ and $\tau_0=1$~ns. The last column is an \emph{estimate} of the size of the
effect, computed from the resistances of Figs.~\ref{fig3} and~\ref{fig13}; the
self-consistent $\Te(x)$ from the loop (\ref{eq:loop}) is what the calculation returns
and is reported in Sec.~\ref{sec:valid}. 

\begin{table*}[t]
\caption{Thermal budget on the $\nu=2$ plateau, computed with the transport solution on the
equilibrium density profile ($d=1.5\,\mu$m, $N=301$, $\TL=1.5$~K): DC longitudinal
resistance, dissipated power per unit length $P/L_y=\int p\,dx=E_y^{0}I$, peak of
$p(x)=j_yE_y$, and the maximum of the self-consistent $\Te(x)$ from Eq.~(\ref{eq:balance})
with the internal-energy relaxation law (\ref{eq:lossU}), $\tau_0=1$~ns, and with
the $T^{5}$ law ($\Sigma=0.1$~W\,m$^{-2}$K$^{-5}$). Model class C of
Table~\ref{tab:models}. The last column is the integrated energy-balance
residual $|\int(\partial_xq_x+P_{\rm loss}-p)\,dx|/\int p\,dx$ of the relaxation solution.
All $\Te$ values are on the cold branch (no back-reaction on $\sigma_l$; see
Sec.~\ref{sec:valid} for the back-reaction). The $98$~pA row is the linear-response
reference of Sec.~\ref{sec:res}.}

\begin{ruledtabular}\begin{tabular}{llllllll}
$B$ (T) & $I$ & $R_L$ ($\Omega$) & $P/L_y$ (W/m) & $p_{\max}$ (W/m$^{2}$) &
$\Te^{\max}$ (K) & $\Te^{\max}-\TL$ (K) & residual \\
 & & & & & Eq.~(\ref{eq:lossU}) & $T^{5}$ law & \\
\hline
$7.035$ & $98$~pA & $1.1\times10^{-6}$ & $3\times10^{-21}$ & $2\times10^{-13}$ & $1.5+10^{-16}$ & $10^{-18}$ & $10^{-15}$ \\
$7.0$ & $1\,\mu$A & $3.5\times10^{-6}$ & $1.2\times10^{-12}$ & $5.2\times10^{-5}$ & $3.35$ & $2\times10^{-5}$ & $4\times10^{-11}$ \\
$7.0$ & $3\,\mu$A & $3.5\times10^{-6}$ & $1.1\times10^{-11}$ & $4.7\times10^{-4}$ & $3.86$ & $1.9\times10^{-4}$ & $8\times10^{-12}$ \\
$7.0$ & $5\,\mu$A & $3.5\times10^{-6}$ & $2.9\times10^{-11}$ & $1.3\times10^{-3}$ & $4.14$ & $5.1\times10^{-4}$ & $2\times10^{-12}$ \\
$6.6$ & $1\,\mu$A & $9.7\times10^{-3}$ & $3.2\times10^{-9}$ & $0.14$ & $5.39$ & $0.050$ & $8\times10^{-16}$ \\
$6.6$ & $3\,\mu$A & $9.7\times10^{-3}$ & $2.9\times10^{-8}$ & $1.2$ & $7.31$ & $0.32$ & $5\times10^{-16}$ \\
$6.6$ & $5\,\mu$A & $9.7\times10^{-3}$ & $8.1\times10^{-8}$ & $3.4$ & $8.44$ & $0.61$ & $2\times10^{-15}$ \\
$6.4$ & $1\,\mu$A & $9.9\times10^{-2}$ & $3.3\times10^{-8}$ & $1.6$ & $6.62$ & $0.37$ & $2\times10^{-15}$ \\
$6.4$ & $3\,\mu$A & $9.9\times10^{-2}$ & $3.0\times10^{-7}$ & $15$ & $10.2$ & $1.23$ & $2\times10^{-16}$ \\
$6.4$ & $5\,\mu$A & $9.9\times10^{-2}$ & $8.2\times10^{-7}$ & $41$ & $12.6$ & $1.83$ & $10^{-17}$ \\
\end{tabular}\end{ruledtabular}
\label{tab:budget}
\end{table*}

Three things follow from the table. At $98$~pA the heating is inactive by sixteen
orders of magnitude, which justifies calling Sec.~\ref{sec:res} linear response in the
thermal as well as the electrical sense. In the microampere range the power depends
enormously on where in the plateau one sits, because $R_L$ does: at the centre
($7.0$~T, $R_L=3.5\,\mu\Omega$) the strip receives $10^{-4}$--$10^{-3}$~W/m$^{2}$, on the
low-field flank ($6.4$~T, $R_L=0.1\,\Omega$) it receives $1$--$40$~W/m$^{2}$. And the two
closures answer differently. With the $T^{5}$ law, which uses a bulk-like electron--phonon
coupling, the centre does not heat at all ($10^{-5}$--$10^{-4}$~K) while the flank heats
by $0.4$--$1.8$~K, comparable to $\TL$. With the relaxation law the strip's activated
heat capacity turns even the centre's $10^{-4}$~W/m$^{2}$ into a $\Te$ of $3.4$--$4.1$~K,
and the flank into $6.6$--$12.6$~K (the $6.4$~T values are not grid-converged, see
Sec.~\ref{sec:valid}). The heating of the narrow low-field strips, which is what
Sec.~\ref{sec:conseq} needs, is therefore present in both closures; the heating of the
broad central strip is a property of the closure, not of the transport solution. Note also that ``$R_{xx}=0$ on the scale of the plot''
and ``finite $R_{xx}$ supplies the Joule power'' are not in contradiction: a resistance
below the plotting resolution still dissipates $R_LI^{2}$, and at microampere currents that
is enough.

\subsection{Consequences}\label{sec:conseq}

$\Te(x)$ enters both Eq.~(\ref{eq:dens}) and Eq.~(\ref{eq:sigl}), and it acts on a strip in
two ways. Through Eq.~(\ref{eq:sigl}) it raises $\sigma_l$ inside the strip by activation, so
the strip loses part of its $1/\rho_l$ weight, current leaks into the compressible regions,
and $R_L$ becomes finite. Through Eq.~(\ref{eq:dens}) it smears the density: a strip of width
$a_k$ spans a fraction $\Delta E_k a_k/|{\rm d}V/{\rm d}x|$ of the gap, and when
$k_B\Te$ becomes comparable to that fraction the flat segment of $\nu(x)$ shrinks and can
fall below $2\lambda$. Both effects act on the narrowest strips first. In the field scan
the narrow strips are those near the edges at low field (Fig.~\ref{fig2}), so the
expectation is that current removes the plateau from its low-field flank first. The
high-field edge is set by $\nu(0)=2$, but heating is not excluded from acting on it:
$\Te$ enters Eq.~(\ref{eq:dens}) and can change the field at which the last
incompressible region survives. Whether that edge moves is therefore a numerical
question, and Sec.~\ref{sec:nl} reports that for the parameters used it does not move
within the resolution of the scan.

\section{Feedback of the current on the landscape}\label{sec:feedback}

In linear response the density profile is the equilibrium one. At finite current the Hall
potential $\mu^{*}(x)$ of Eq.~(\ref{eq:RHRL}) adds to the confinement in Eq.~(\ref{eq:dens}),
the density adjusts, the strips move and change width, and $\shat$ must be recomputed. The
closed loop is
\begin{equation}
\nu(x)\xrightarrow{(\ref{eq:sigl}),(\ref{eq:RHRL})}
\{j_y,\mu^{*},p\}\xrightarrow{(\ref{eq:local})}\Te(x)
\xrightarrow{(\ref{eq:Vneutral}),(\ref{eq:dens})}\nu(x),
\label{eq:loop}
\end{equation}
iterated at fixed $I$. Because $\mu^{*}(x)$ is odd about the centre of the bar while
$V_{\rm bg}$ and the kernel are even, the driven boundary condition breaks the reflection
symmetry of the problem (this is an explicit, not a spontaneous, symmetry breaking): the effective confinement is steepened at one edge and flattened at the
other, the strip at the edge of the \emph{higher} electrochemical potential widens and the
other narrows \cite{Siddiki2009EPL87,Gerhardts2013}. Under $I\to-I$ the exact solution satisfies $\nu(x;I)=\nu(-x;-I)$, and likewise for
$\Te$, $j_y$, $\mu^{*}$ and $p$; the size of the residual of this relation in the
numerical solution is the diagnostic that separates the physical asymmetry from a bias of
the continuation, and it is reported in Sec.~\ref{sec:valid}.
Heating inherits the asymmetry through Eq.~(\ref{eq:joule}): the wider strip carries the
larger share of the current and therefore of the power.

\section{The AC local Ohm's law}\label{sec:ac}

\subsection{Linear response about the DC profile}\label{sec:lin}

Write all fields as $X(x,t)=X_0(x)+\mathrm{Re}[\delta X(x,\omega)e^{-i\omega t}]$, with $X_0$
the converged DC solution, and keep first order in $\delta X$. The DC profile fixes the local
conductivities and compressibility once and for all, so the frequency sweep is a linear
problem on top of it (frozen-profile assumption). The one structural change is that
$\nabla\cdot\jvec$ no longer vanishes: with $\varrho=-e\,\delta n$,
\begin{equation}
\partial_x\,\delta j_x(x,\omega)=-i\omega e\,\delta n(x,\omega),
\label{eq:cont}
\end{equation}
so a transverse current may terminate on accumulated charge --- the microscopic content of a
displacement current, with nothing added by hand. Retardation is negligible for
$\omega\ll c/(\sqrt{\kappa}\,d)$; linearising Eqs.~(\ref{eq:Vneutral}) and (\ref{eq:dens}),
\begin{equation}
\begin{gathered}
\delta n=D_T\left[\delta\eta-\delta V\right],\qquad \delta V=\hat K\delta n,\\
\delta n=\hat{M}\,\delta\eta,\qquad \hat{M}=\left[1+\hat{D}_T\hat{K}\right]^{-1}\hat{D}_T,
\end{gathered}
\label{eq:M}
\end{equation}
where $D_T=\partial n_{\rm el}/\partial\mu$ is the local thermodynamic density of states and
$\delta\eta$ the local shift of the electrochemical potential. In compressible regions $D_T$
is large and the electrons screen; inside a strip $D_T\to0$ and they cannot. That contrast
generates the whole frequency dependence found below.

\subsection{The dynamic conductivity tensor, and why it does not matter}\label{sec:sigac}

In a relaxation-time description carrier inertia is a purely inductive correction,
$\rhohat(\nu,\omega)=\rhohat^{\rm DC}(\nu)-i\omega(m^{*}/n_{\rm el}e^{2})\hat 1$, of relative
order $\varepsilon_\omega=\omega/\omega_c$ against the Hall resistivity. Far below cyclotron
resonance the fast subsystem follows the drive adiabatically and the local tensor may be
taken quasistatic. Disorder-assisted absorption, collective edge modes \cite{EMP} and
contact or circuit admittances are outside the model; Eqs.~(\ref{eq:cont}) and (\ref{eq:M})
isolate the electrostatic contribution. Thermally, the time-averaged Joule power at finite
frequency is $\langle p\rangle=\tfrac12\mathrm{Re}(\delta\jvec\cdot\delta\Evec^{*})$: the
reactive part of the response, which is all that changes below the megahertz range, carries
no dissipation and does not heat the strips. This is why $\mathrm{Im}\,R_L$ must not be
called dissipation, and why the thermal budget of Sec.~\ref{sec:budget} uses
$\mathrm{Re}\,R_L$, not $|R_L|$.

\subsection{Closed equation and the measured resistances}

With $E_x=(1/e)\partial_x\delta\eta$ and $E_y$ the imposed uniform drive,
$j_x=\sigma_lE_x+\sigma_HE_y$, $j_y=-\sigma_HE_x+\sigma_lE_y$, and
\begin{equation}
\begin{gathered}
\left[\frac{1}{e}\partial_x\!\left(\sigma_l\,\partial_x\delta\eta\right)
+i\omega e\,\hat{M}\delta\eta\right](x)=-E_y\,\partial_x\sigma_H(x),\\
j_x(\pm d)=0 .
\end{gathered}
\label{eq:closed}
\end{equation}
At $\omega=0$ this integrates to $j_y=E_y/\rho_l$, recovering Eq.~(\ref{eq:RHRL}). The
measured quantities generalise to the complex
\begin{equation}
R_H(\omega)=-\frac{\int E_x\,dx}{\int j_y\,dx},\qquad
R_L(\omega)=\frac{2d\,E_y}{\int j_y\,dx}.
\label{eq:RHac}
\end{equation}
We use the gradient of the electrochemical potential throughout; using the electric field
instead differs by a diffusive term that is negligible wherever $D_T$ is large ---
everywhere except in the strips, which control the response.

\section{Numerical scheme}\label{sec:num}

Equations~(\ref{eq:Vneutral}), (\ref{eq:dens}) and (\ref{eq:closed}) are discretised on
$N=1001$ cell centres, with the logarithmic singularity of the kernel integrated analytically
over the diagonal cell; the sum rule is satisfied to $10^{-4}$. The $B=0$, $T\to0$ state is
obtained by shooting on $\mu^{*}$ until the channel half width reaches $b/d=0.9$, which
fixes $E_F^{0}$; $\mu^{*}$ is then held fixed as $B$ is varied. Each field point is solved
by damped Newton iteration on the residual $N(\mu^{*}-V[n])-n$, converged to
$2\times10^{-6}$ in units of $n_0$. On the converged profile the AC equation is assembled in
flux form and solved as a complex linear system at each frequency; a sweep of $85$
frequencies from $1$~Hz to $1$~THz costs little more than a DC run. For the finite-current runs
of Sec.~\ref{sec:nl} (class B) the electrostatic part of the loop (\ref{eq:loop}) is
iterated at fixed $I$ with $\Te=\TL$: the Ohm law is solved on the current profile,
$\mu^{*}(x)$ is inserted into Eq.~(\ref{eq:dens}), and the Newton step is repeated until
the density and the potential are converged.
Near $B_c$, where $\rho_l$ inside the current-carrying strip varies by six orders of
magnitude across a few grid cells, the Jacobian is stiff and the damped Newton step is
limited by backtracking; the field scan is continued from the preceding solution, and the
dependence of the converged state on that continuation is one of the checks listed in
Sec.~\ref{sec:valid}. The thermal equation (\ref{eq:balance}) is solved on the converged transport
solution by a damped Newton iteration in $\Te(x)$, started from the conduction-free
local balance obtained by bisection in $\ln\Te$ cell by cell (this starting guess is
needed because $c_{\rm e}$ inside a strip spans many decades). Convergence is monitored by
the local and integrated residuals of Eq.~(\ref{eq:balance}) and by the identity
$\int p\,dx=E_y^{0}I$, which holds to machine precision; the integrated residual is below
$10^{-10}$ of $\int p\,dx$ at every point of Table~\ref{tab:budget}. The thermal
solutions of Sec.~\ref{sec:valid} are computed on a $d=1.5\,\mu$m bar with $N=301$
cells, on the equilibrium density profile, and the back-reaction of $\Te$ is carried
through $\sigma_l(\Te)$ only; the reflection test of Sec.~\ref{sec:valid} uses the
self-consistent finite-current code on the $d=2.5\,\mu$m bar of Fig.~\ref{fig9} with
$N=750$.
The results are reported in units of $d$ and $\hbar\omega_c$ so that runs at different
fields can be overlaid.

Four levels of coupling occur in this paper and are kept apart; Table~\ref{tab:models}
names them and every figure caption states which one it shows.

\begin{table*}[t]
\caption{Levels of coupling between electrostatics, transport and temperature. Class D,
the fully coupled calculation, is the one that would describe the hot branch and the
thermal contribution to the plateau narrowing; it has not been done here.}

\begin{ruledtabular}\begin{tabular}{lp{3.2cm}p{4.6cm}p{4.6cm}}
Class & Hall-potential feedback on $n$ & $\Te$ & Where \\
\hline
A & no & $\TL$ & Figs.~\ref{fig1}--\ref{fig5} (equilibrium profile, linear response) \\
B & yes & $\TL$ & Figs.~\ref{fig6}--\ref{fig13}, Fig.~\ref{fig15}(a) \\
C & no & post-processed on the class-A profile & Table~\ref{tab:budget}, Fig.~\ref{fig14} \\
C$'$ & no & back-reaction through $\sigma_l(\Te)$ only & Fig.~\ref{fig15}(b,c) \\
D & yes & jointly converged with $n$, $V$, $\mu^{*}$ & not performed \\
\end{tabular}\end{ruledtabular}
\label{tab:models}
\end{table*}

Two reference states occur in this paper and should be kept apart. Figures~\ref{fig1},
\ref{fig2}, \ref{fig3} and~\ref{fig5} share $\nu(0)B=14.76$~T, so $\nu(0)=2$ at
$7.38$--$7.42$~T (the small difference is the field step of the scan). Figures~\ref{fig4}
and \ref{fig6}--\ref{fig13} share $\nu(0)B=15.2$~T, so $\nu(0)=2$ at $7.6$~T. Within each
family the comparisons are exact; across families, positions in field should be compared
through $\nu(0)$, which the figures carry.

Sample parameters are $d=1.5\,\mu$m, $n_0=4\times10^{11}$~cm$^{-2}$, $b/d=0.9$,
$\TL=1.5$~K, $\lambda=30$~nm and $\Gamma\propto\sqrt{B}$ with $\gamma_I=0.1$ at $10$~T.
They are chosen to coincide with Ref.~\cite{SG2004}, so that the linear-response results
can be compared with the published screening-theory calculation, and they are
representative of gate-defined GaAs/AlGaAs bars of a few micrometres width; higher
mobility (smaller $\Gamma$) narrows the strips and lowers $\sigma_l$ inside them, which
moves the low-field edge and the thermal budget in a direction that is calculable from
Eqs.~(\ref{eq:sigl}) and (\ref{eq:joule}) but is not scanned here. The
$B=0$ reference gives $n_{\rm el}(0)=3.578\times10^{11}$~cm$^{-2}$ and $E_F^{0}=12.784$~meV,
against $3.55\times10^{11}$ and $12.69$~meV in Ref.~\cite{SG2004}.

\section{Results I: linear response}\label{sec:res}

In the figures the local longitudinal conductivity is labelled $\sigma_{xx}$ or
$\sigma_l$; they are the same quantity. Figures~\ref{fig1}--\ref{fig3} give the DC solution
in linear response, Figs.~\ref{fig4} and~\ref{fig5} repeat the profiles and the maps at
finite frequency on the same scales, and Fig.~\ref{fig6} carries the resistance scan to a
moderate current at $1$~kHz.

\subsection{DC profiles across the bar}\label{sec:profsym}

\begin{figure*}[t]\centering
\includegraphics[width=\textwidth]{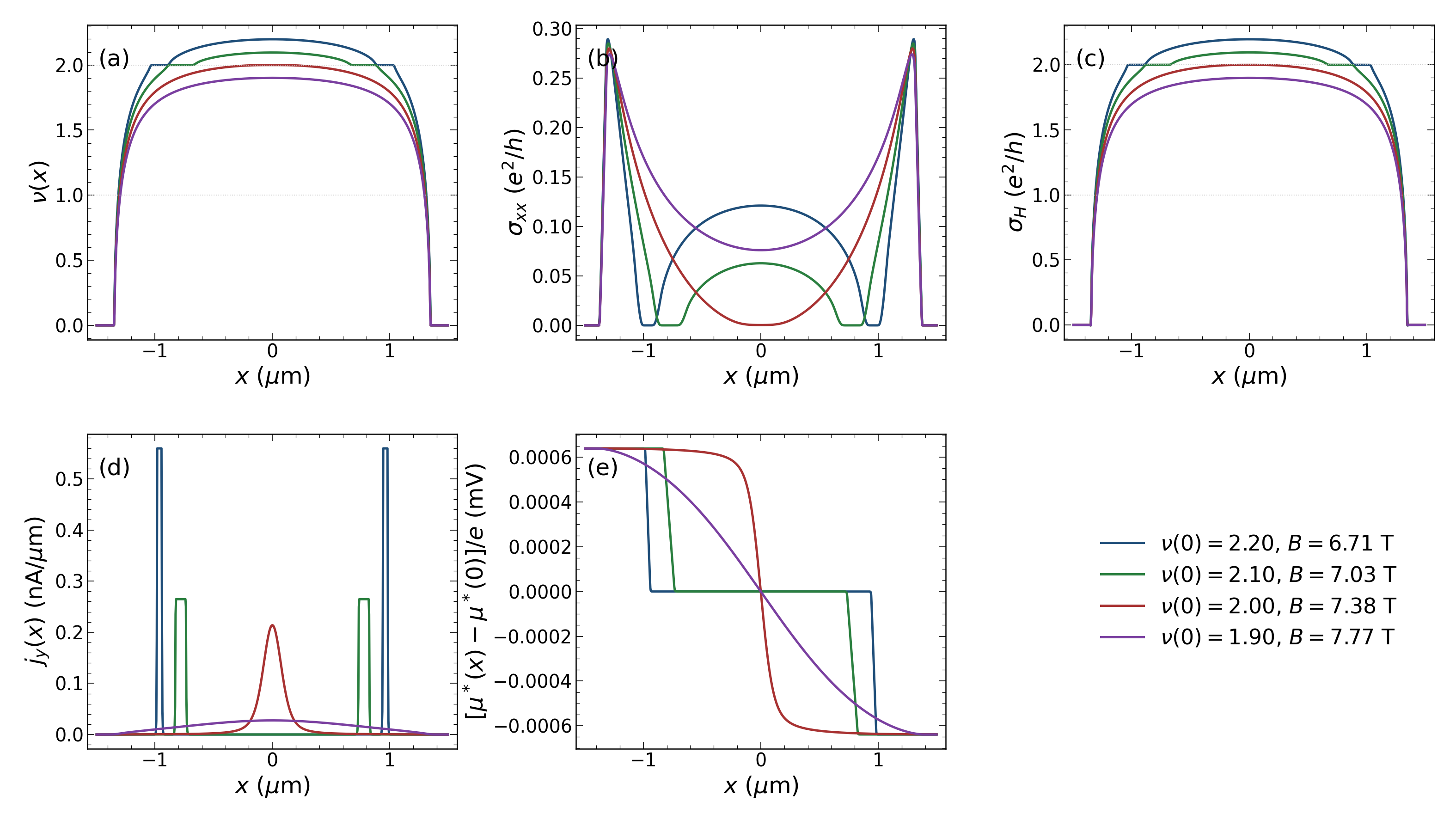}
\caption{Class A. DC solution in linear response ($I\simeq98$~pA, $eV_H/\hbar\omega_c\simeq10^{-4}$)
at the four fields in the legend, labelled by the central filling factor. (a) Local filling
factor; the flat segments at $\nu=2$ are the incompressible strips. (b) Local longitudinal
conductivity, whose zeros locate the strips. (c) Local Hall conductivity, tracking $\nu(x)$.
(d) Current density, confined to the strips and collapsing onto a central peak at
$\nu(0)=2.00$; by Eq.~(\ref{eq:joule}) this is also the profile of the Joule power.
(e) Electrochemical potential relative to the centre of the bar, antisymmetric, with steps
where the current density peaks.}
\label{fig1}
\end{figure*}

Figure~\ref{fig1} shows the converged solution at four fields spanning the $\nu=2$ plateau,
labelled by the central filling factor: $2.20$ at $6.71$~T, $2.10$ at $7.03$~T, $2.00$ at
$7.38$~T and $1.90$ at $7.77$~T. The products $\nu(0)B$ are constant to four digits
($14.76$~T), the simplest check that $\mu^{*}$ is genuinely held fixed. Panel~(e) spans
$\pm0.63\,\mu$V, so $V_H=1.26\,\mu$V, $I=V_H/R_H\simeq98$~pA and
$eV_H/\hbar\omega_c\simeq10^{-4}$. The solution is symmetric under $x\to-x$: the strips sit
at $\pm0.98\,\mu$m at $\nu(0)=2.20$ and $\pm0.79\,\mu$m at $2.10$, and the Hall voltage
divides equally between them, as Eq.~(\ref{eq:weight}) requires of an even $\rho_l$.

The sequence displays the whole mechanism. At $\nu(0)=2.20$ the flat segments of $\nu(x)$
sit near the edges and the zeros of $\sigma_{xx}$ are narrow; as the field rises the strips
move inwards and broaden until at $\nu(0)=2.00$ the incompressible region has swallowed the
bulk; at $1.90$ no incompressible region exists. Panels~(d) and~(e) are the transport
counterparts: the current is confined to the strips, peaking at $0.56$~nA/$\mu$m at
$\nu(0)=2.20$ and $0.27$~nA/$\mu$m at $2.10$, collecting into a single central peak at
$\nu(0)=2.00$ where $\sigma_l$ is smallest, and the electrochemical potential steps exactly
where the current peaks. By Eq.~(\ref{eq:joule}) panel~(d) is also the map of the Joule
power, and at $98$~pA that power is $2\times10^{-21}$~W per metre of bar, which by
Table~\ref{tab:budget} raises $\Te$ inside a strip by $10^{-16}$~K: linear response in the
thermal as well as the electrical sense. The three types of Hall-potential profile seen by
scanning force microscopy \cite{Weitz2000a,Ahlswede1,Ahlswede2} --- step-like, single-drop
and smooth --- appear without further input.

\subsection{The same quantities along the field scan}\label{sec:dcmaps}

\begin{figure*}[t]\centering
\includegraphics[width=\textwidth]{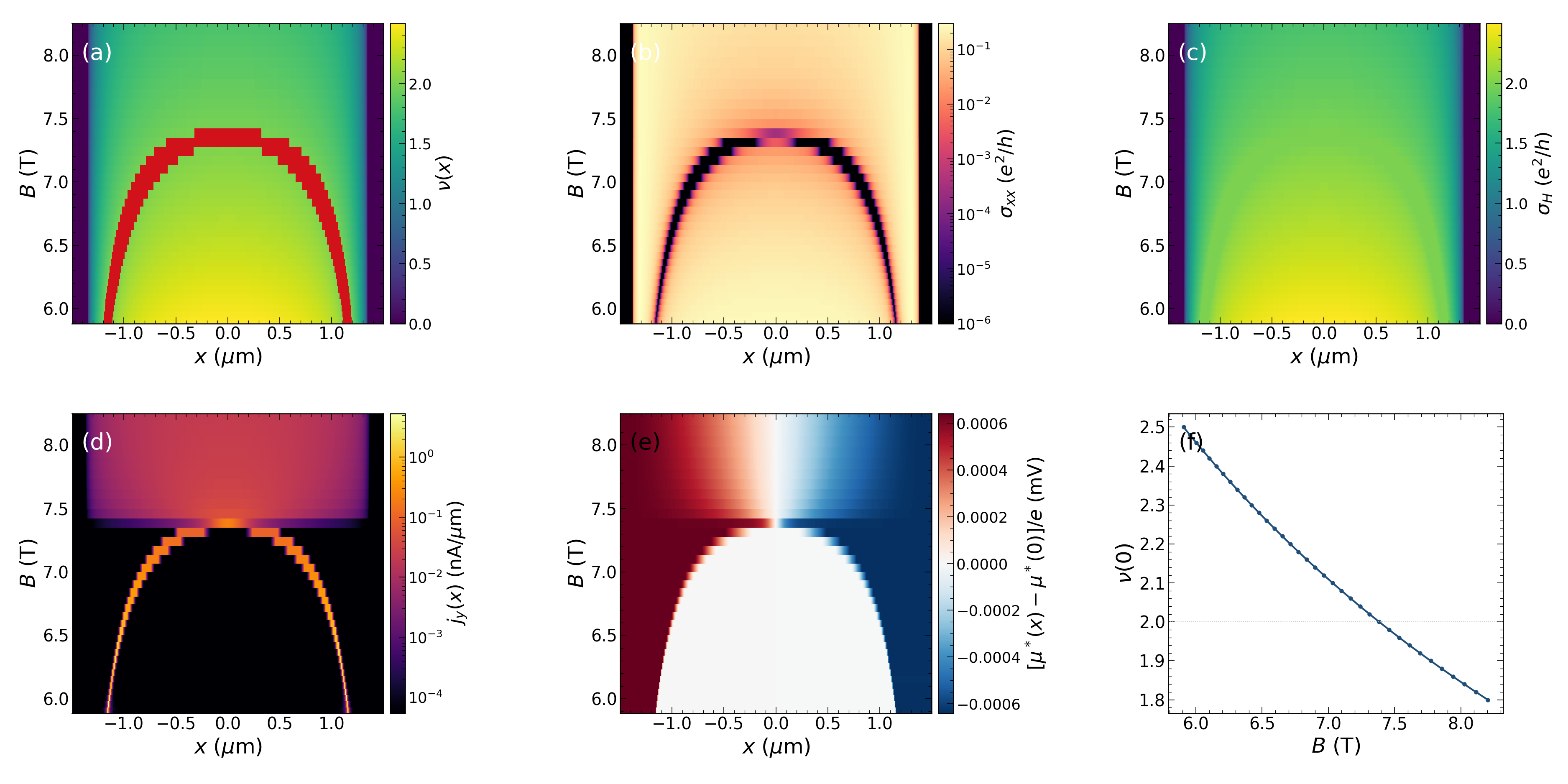}
\caption{Class A. The quantities of Fig.~\ref{fig1} along the field scan, in the same
linear-response limit. (a) Local filling factor with the incompressible region $\nu=2$ in
red: a band of finite width that migrates outwards as the field is lowered and closes across
the centre between $7.28$ and $7.42$~T. (b) $\sigma_{xx}$ on a logarithmic scale, five
decades below the compressible background along the same crescent. (c) $\sigma_H$.
(d) Current density on a logarithmic scale, following the arcs; above $7.6$~T it spreads
across the bar. (e) Electrochemical potential relative to the centre, the step tracking the
strips. (f) Central filling factor, crossing two at $7.42$~T. The solution is symmetric
under $x\to-x$ throughout.}
\label{fig2}
\end{figure*}

Figure~\ref{fig2} assembles the five quantities as maps in the $(x,B)$ plane. The red region
in panel~(a) is where $\nu=2$; it is a band of finite width --- the strip width --- that
closes across the centre between $7.28$ and $7.42$~T and opens towards the edges as the
field is lowered, the crescent of the screening literature \cite{Guven2003,SG2004}. The
region and its mirror image agree over $99.2\%$ of their area. Panel~(b) shows
$\sigma_{xx}$ falling by five decades along that crescent and nowhere else; panel~(d) shows
the current following it. Above the plateau the current spreads across the bar. Panel~(e)
shows the step in the Hall potential tracking the strip position, the observation that made
the two descriptions of the effect empirically distinct \cite{Ahlswede1,Ahlswede2}.
Panel~(f) fixes the field at which $\nu(0)$ passes through two, $7.42$~T. The fine
horizontal striping in (b), (d), (e) is the granularity of the field scan, not structure.

\subsection{Quantised resistances, and why the two plateau edges differ}\label{sec:edges}

\begin{figure}[t]\centering
\includegraphics[width=\columnwidth]{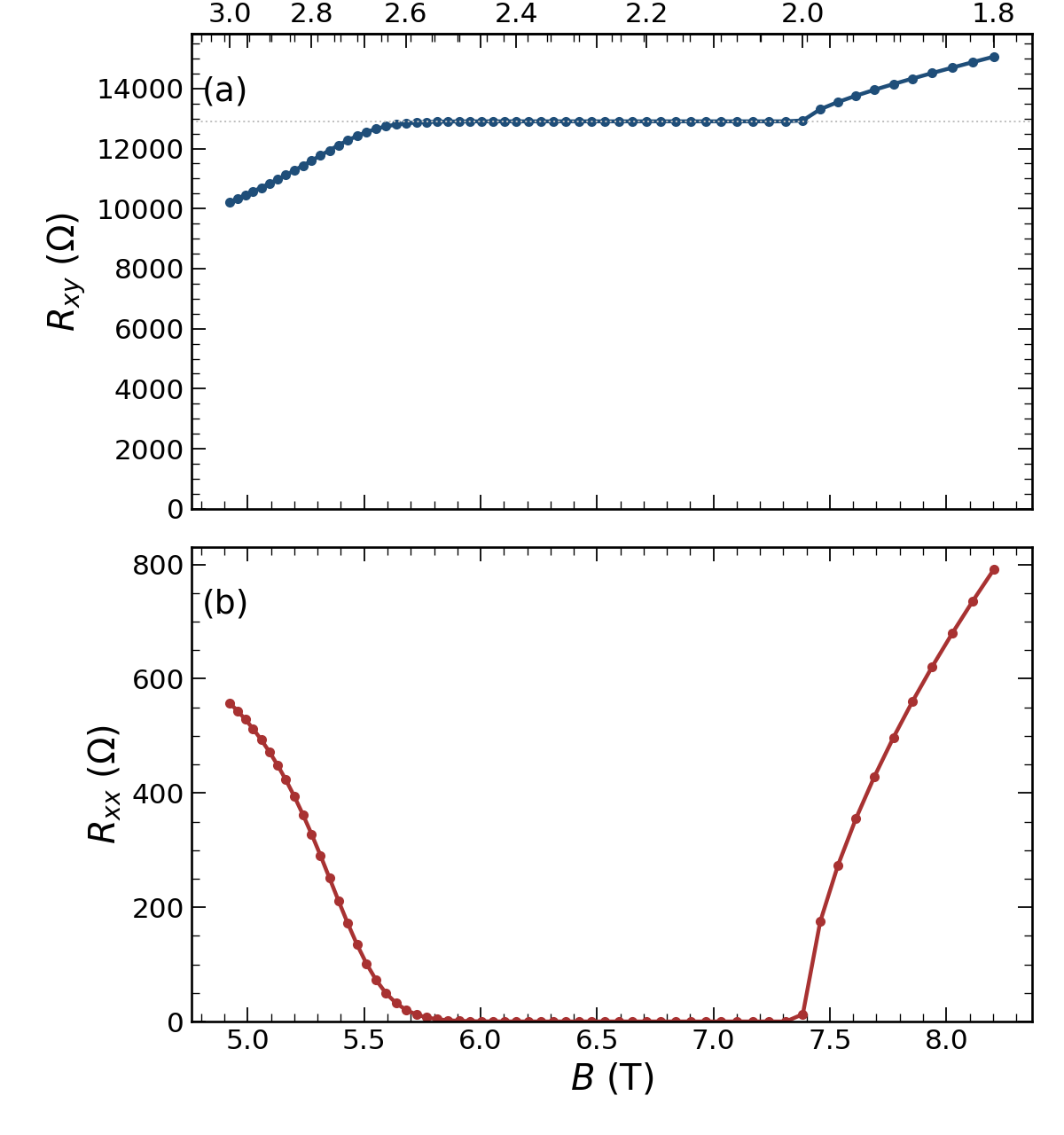}
\caption{Class A. (a) Hall and (b) longitudinal resistance from Eq.~(\ref{eq:RHRL}) along the scan of
Fig.~\ref{fig2}; the upper axis gives $\nu(0)$ and the dotted line $h/2e^{2}$. The plateau
runs from $5.7$~T to $7.42$~T. The high-field edge falls exactly at $\nu(0)=2$, where the
last incompressible region disappears; the low-field edge is set by the removal of strips
narrower than $2\lambda$ and is kernel-dependent.}
\label{fig3}
\end{figure}

Figure~\ref{fig3} gives the resistances of Eq.~(\ref{eq:RHRL}) along the scan. $R_{xy}$ is
pinned at $h/2e^{2}=12\,906.4\,\Omega$ from about $5.7$~T to $7.42$~T, and $R_{xx}$ vanishes
on the scale of the plot over the same interval, i.e.\ $2.00\lesssim\nu(0)\lesssim2.6$
on the upper axis. The plateau coincides with the interval in which a strip wider than
$2\lambda$ exists somewhere in the bar.

The two edges are not alike. On the high-field side the plateau ends abruptly at
$\nu(0)=2.00$: once the central filling factor falls below two no region can have $\nu=2$,
the weight $1/\rho_l$ loses its divergence everywhere at once and $R_{xx}$ rises
immediately. That edge is a property of the electrostatics alone. On the low-field side the
plateau ends gradually, around $\nu(0)\simeq2.6$, because the strips narrow and are removed
one by one as their width falls below $2\lambda$; that edge is kernel-dependent
\cite{SG2004b}. The plateau therefore sits entirely on the low-field side of the field at
which $\nu(0)=2$ --- one-sided, not merely asymmetric --- which is the interaction-mediated
plateau asymmetry reported for gate-defined bars, where the classical Hall line cuts the
even-integer plateaus off centre towards larger $B$ \cite{Siddiki2009EPL88}. The
prediction, sharpened in Sec.~\ref{sec:conseq}, is that anything that weakens the narrow
strips --- the kernel, temperature, or current --- acts on the low-field flank only. The
plateau width agrees with Ref.~\cite{SG2004} ($\nu(0)$ up to $2.49$) to within the
resolution of the scan, the high-field edge exactly.

\subsection{AC in linear response}\label{sec:aclin}

\emph{Working point.} Fixing $\nu(0)=2.1$ gives $B=7.035$~T, $\hbar\omega_c/E_F^{0}=0.951$
and $\Gamma=1.449$~meV. The strips occupy $0.653<|x|<0.902\,\mu$m and are $249$~nm wide;
the averaged $\sigma_l$ falls to $5.8\times10^{-13}\,e^{2}/h$ inside them
($4.4\times10^{-17}$ before averaging). At DC, $R_H=0.4999359\,h/e^{2}$ and
$R_L=2.27\times10^{-11}\,h/e^{2}$, with more than $99.99\%$ of the current in the strips.

\begin{figure*}[t]\centering
\includegraphics[width=\textwidth]{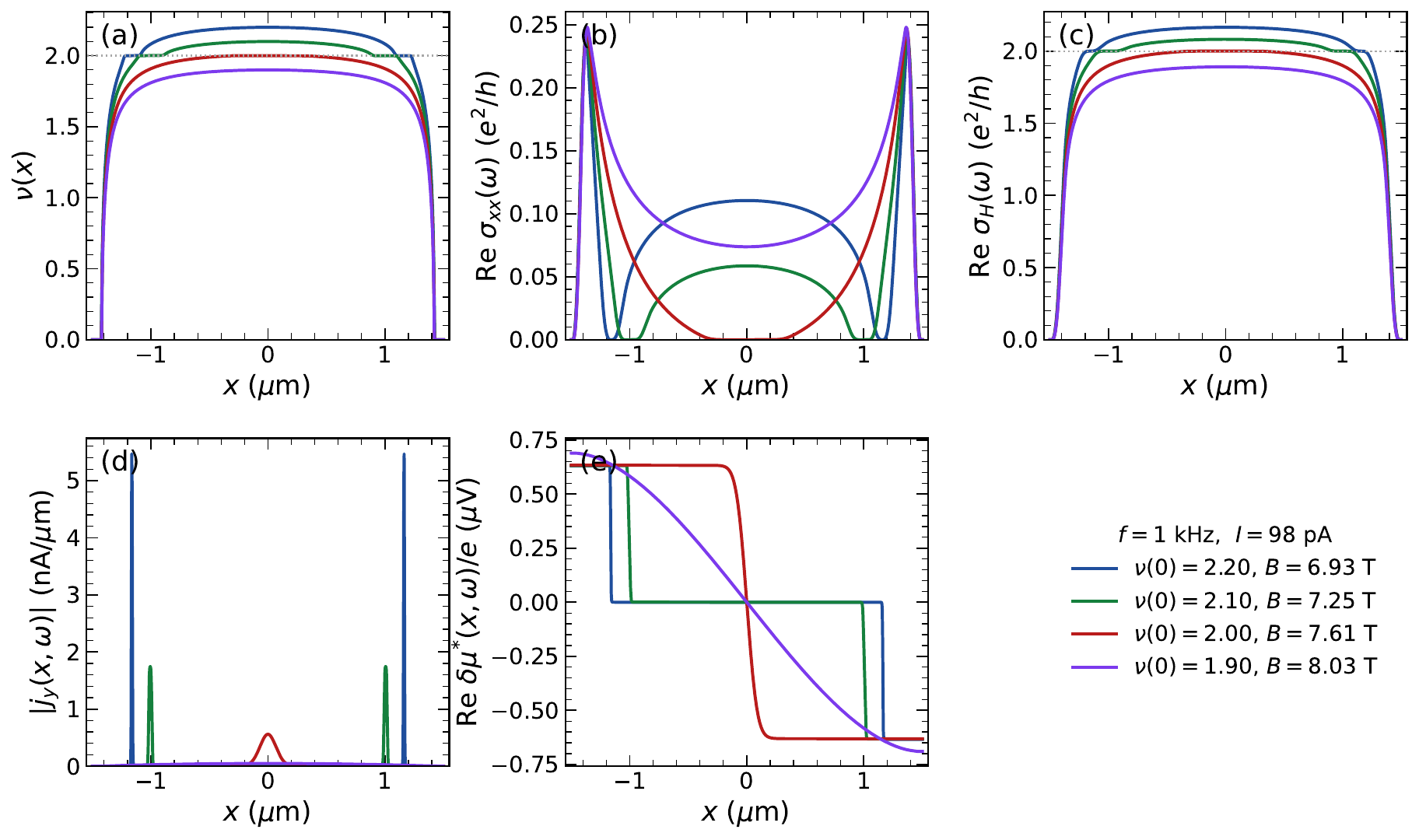}
\caption{Class A. As Fig.~\ref{fig1}, for the AC solution of Eq.~(\ref{eq:closed}) at $f=1$~kHz and
$I=98$~pA, at the four central filling factors in the legend. (a) $\nu(x)$;
(b) $\mathrm{Re}\,\sigma_{xx}(\omega)$; (c) $\mathrm{Re}\,\sigma_H(\omega)$;
(d) $|j_y(x,\omega)|$, confined to the strips; (e) $\mathrm{Re}\,\delta\mu^{*}(x,\omega)/e$,
with steps where the current peaks. At this frequency the AC and DC solutions have the same
structure; the differences are those of Table~\ref{tab:res}.}
\label{fig4}
\end{figure*}

\emph{Profiles at $1$~kHz.} Figure~\ref{fig4} shows the AC solution of
Eq.~(\ref{eq:closed}) at $f=1$~kHz and $I=98$~pA, with the same five observables as
Fig.~\ref{fig1}, at $\nu(0)=2.20$, $2.10$, $2.00$ and $1.90$. Panel for panel it is
Fig.~\ref{fig1}: the strips are flat segments of $\nu(x)$, $\mathrm{Re}\,\sigma_{xx}$ falls
into deep minima there, $\mathrm{Re}\,\sigma_H$ follows $\nu(x)$, $|j_y|$ is confined to the
strips and collapses to the centre as $\nu(0)\to2$, and $\mathrm{Re}\,\delta\mu^{*}$ is flat
in the compressible regions with steps at the strips. The fields at which a given $\nu(0)$
occurs differ slightly from Fig.~\ref{fig1} ($\nu(0)B=15.2$~T) because this run uses the
reference state of Figs.~\ref{fig10}--\ref{fig13}. Nothing in the landscape responds to the
drive: $\varepsilon_\omega=3\times10^{-10}$ at $1$~kHz, and the electrostatic term
$i\omega e\hat M$ is ten orders below the $\sigma_{xx}$-weighted term outside the strips.
Between DC and $1$~GHz the relative deviation of $|j_y(x)|$ from the DC profile grows
strictly linearly in $\omega$ from $10^{-12}$ at $1$~Hz; the share of current carried by
the strips is $1.000000$ through $1$~MHz and $0.993$ at $1$~GHz.

\emph{Frequency dependence of the resistances.} The solution obeys
\begin{equation}
\begin{gathered}
\mathrm{Im}\,R_H\propto\omega,\qquad |R_L-R_L^{\rm DC}|\propto\omega,\\
\mathrm{Re}\,R_H-R_H^{\rm DC}\propto\omega^{2},
\end{gathered}
\label{eq:powerlaws}
\end{equation}
with fitted exponents $1.000$ and $1.999$ (Table~\ref{tab:res}). Causality requires
reactive terms to be odd and real corrections even in $\omega$, which is why the modulus of
$R_L$ must not be called dissipation. The change in the real Hall resistance stays below
$10^{-9}$ in relative terms until $1.9$~MHz, below $10^{-6}$ until $72$~MHz and below
$10^{-3}$ until $1.9$~GHz. The intrinsic electrostatic correction calculated within the frozen-profile quasilocal
model is therefore considerably smaller than typical uncompensated AC-QHR deviations
\cite{ACQHE}. Edge magnetoplasmons \cite{EMP}, disorder-assisted absorption and
thermoelectric dynamics are outside the model, so this bounds one intrinsic contribution
rather than locating the origin of the observed deviations. Measured AC Hall resistances typically show a \emph{linear} frequency dependence
that is attributed to device and dielectric losses and to the measurement circuit
\cite{ACQHE,Kalmbach2014}; the present model, which contains none of these, gives a
reactive first-order term and a real correction of second order. The two are therefore
not in conflict, but neither is the model a calculation of the measured quantity: it
bounds one intrinsic contribution, at $10^{-9}$ in $\mathrm{Im}\,R_H$ at kilohertz.

\begin{table*}[t]
\caption{Complex Hall resistance and magnitude of the longitudinal impedance at
$B=7.035$~T, $\nu(0)=2.1$. Only $|R_L|$ is tabulated; the dissipative part
$\mathrm{Re}\,R_L$, which is what enters the thermal budget, equals its DC value to the
precision shown below $1$~MHz and must be exported separately at higher frequencies.}

\begin{ruledtabular}\begin{tabular}{lrrr}
$f$ & $\mathrm{Re}\,R_H$ ($h/e^{2}$) & $\mathrm{Im}\,R_H$ ($h/e^{2}$) & $|R_L|$ ($h/e^{2}$) \\
\hline
DC       & $0.499935949$ & $0$                   & $2.27\times10^{-11}$ \\
$1$~Hz   & $0.499935949$ & $-1.85\times10^{-12}$ & $3.46\times10^{-11}$ \\
$1$~kHz  & $0.499935949$ & $-1.82\times10^{-9}$  & $2.23\times10^{-8}$ \\
$1$~MHz  & $0.499935949$ & $-1.82\times10^{-6}$  & $2.23\times10^{-5}$ \\
$1$~GHz  & $0.500078868$ & $-1.81\times10^{-3}$  & $2.22\times10^{-2}$ \\
\end{tabular}\end{ruledtabular}
\label{tab:res}
\end{table*}

\begin{figure*}[t]\centering
\includegraphics[width=\textwidth]{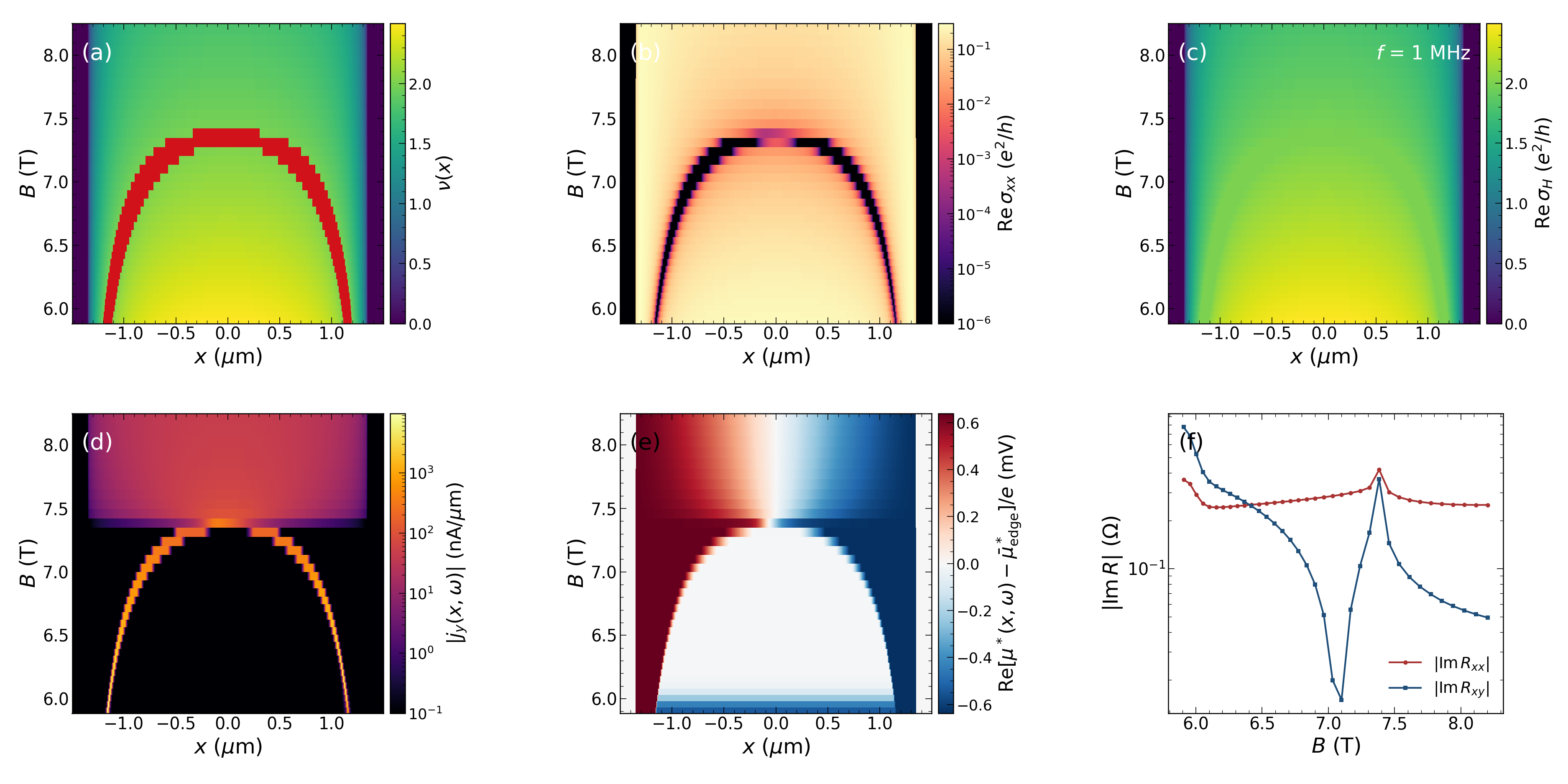}
\caption{Class A. As Fig.~\ref{fig2}, for the AC solution at $f=1$~MHz. (a) $\nu(x)$ with the
$\nu=2$ region in red. (b) $\mathrm{Re}\,\sigma_{xx}$. (c) $\mathrm{Re}\,\sigma_H$.
(d) $|j_y(x,\omega)|$. (e) $\mathrm{Re}[\mu^{*}(x,\omega)-\bar\mu^{*}_{\rm edge}]/e$.
(f) $|\mathrm{Im}\,R_{xx}|$ and $|\mathrm{Im}\,R_{xy}|$ along the scan; the real parts are
indistinguishable from Fig.~\ref{fig3}. The whole difference between AC and DC is
panel~(f), of order $10^{-1}\,\Omega$.}
\label{fig5}
\end{figure*}

\emph{The field scan at $1$~MHz.} Figure~\ref{fig5} is Fig.~\ref{fig2} recomputed with the
complex equation at $f=1$~MHz. Panels (a)--(e) cannot be told from their DC counterparts:
same red band, same crescent in $\mathrm{Re}\,\sigma_{xx}$, same arcs of current, same step
in the potential. Panel~(f), on its own logarithmic axis, is where the whole AC content
lives: $|\mathrm{Im}\,R_{xx}|$ and $|\mathrm{Im}\,R_{xy}|$ are of order $10^{-1}\,\Omega$,
$10^{-5}$ of the quantised value, consistent with Table~\ref{tab:res}. Two features carry
information: $|\mathrm{Im}\,R_{xy}|$ changes sign at $7.1$~T, and both reactive parts peak
sharply at the high-field edge, $7.4$~T, where the strips are narrow and $\sigma_l$ inside
them is largest. If the intrinsic AC response is ever to be measured, the place to look is
the plateau edge.

\subsection{Resistances at moderate current and finite frequency}\label{sec:fig6}

\begin{figure}[t]\centering
\includegraphics[width=\columnwidth]{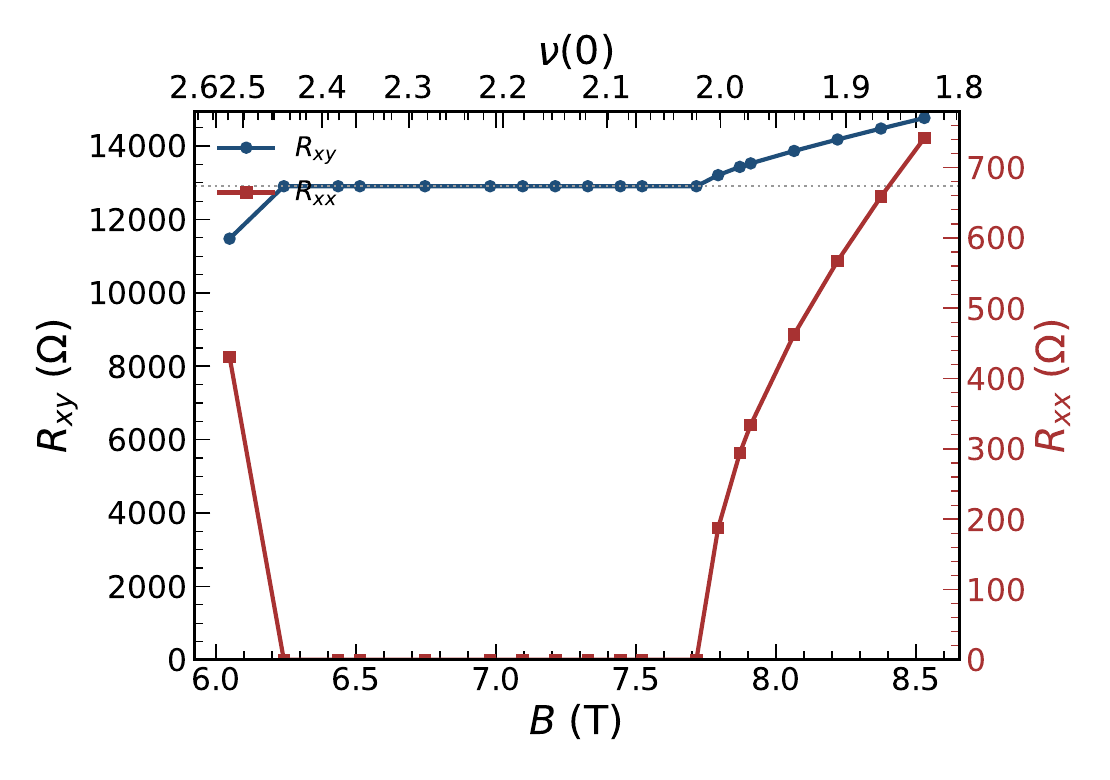}
\caption{Class B. Hall (left axis) and longitudinal (right axis) resistance versus field at
$f=1$~kHz about a DC profile computed at $I=408$~nA, $eV_H/\hbar\omega_c=0.425$. The upper
axis gives $\nu(0)$ and the dotted line $h/2e^{2}$. The plateau runs from $6.25$ to
$7.72$~T, $2.00\lesssim\nu(0)\lesssim2.45$: the quantised value and the high-field edge are
unchanged from Fig.~\ref{fig3}, and the plateau is narrower on the low-field flank only.}
\label{fig6}
\end{figure}

Figure~\ref{fig6} is the resistance scan computed at $f=1$~kHz about a DC profile
determined self-consistently at $I=408$~nA ($V_H=5.366$~mV, $eV_H/\hbar\omega_c=0.425$),
the first point at which the feedback of Sec.~\ref{sec:feedback} is active. $R_{xy}$ is
pinned at $h/2e^{2}$ from $6.25$ to $7.72$~T and $R_{xx}$ vanishes on the scale of the plot
over the same interval: the quantisation survives a drive four decades above
Fig.~\ref{fig3} and survives it at the same value. What changes is the width, from
$2.00\lesssim\nu(0)\lesssim2.6$ to $2.00\lesssim\nu(0)\lesssim2.45$, and the whole loss
sits on the low-field flank; the high-field edge, fixed by $\nu(0)=2$, is unmoved. This is
the first instance of the one-sided narrowing predicted in Sec.~\ref{sec:conseq}. Because
the frequency is low enough that $\hat\sigma(\omega)$ cannot be responsible
(Sec.~\ref{sec:sigac}), whatever separates Fig.~\ref{fig6} from Fig.~\ref{fig3} is the
amplitude, not the frequency.

\section{Results II: finite current, broken symmetry and local heating}\label{sec:nl}

The runs of this section impose $I=1$, $3$ and $5\,\mu$A on the same bar and iterate the
loop (\ref{eq:loop}) to convergence. Since $R_H=h/2e^{2}$ on the plateau the Hall voltages
are $12.9$, $38.7$ and $64.5$~mV, i.e.\ $eV_H/\hbar\omega_c\simeq1$, $3$ and $5$ at $7$~T:
the Hall drop is one to five cyclotron gaps, and the current densities $I/2d=0.33$--$1.7$~A/m
bracket the experimental breakdown range of Table~\ref{tab:scales}. Positions are given in
units of $d$, potentials in units of $\hbar\omega_c$, and $j_y$ in units of $I/d$, so that
runs at different fields and currents overlay.

\subsection{Profiles at three fields and three currents}\label{sec:fig7}

\begin{figure*}[t]\centering
\includegraphics[width=\textwidth]{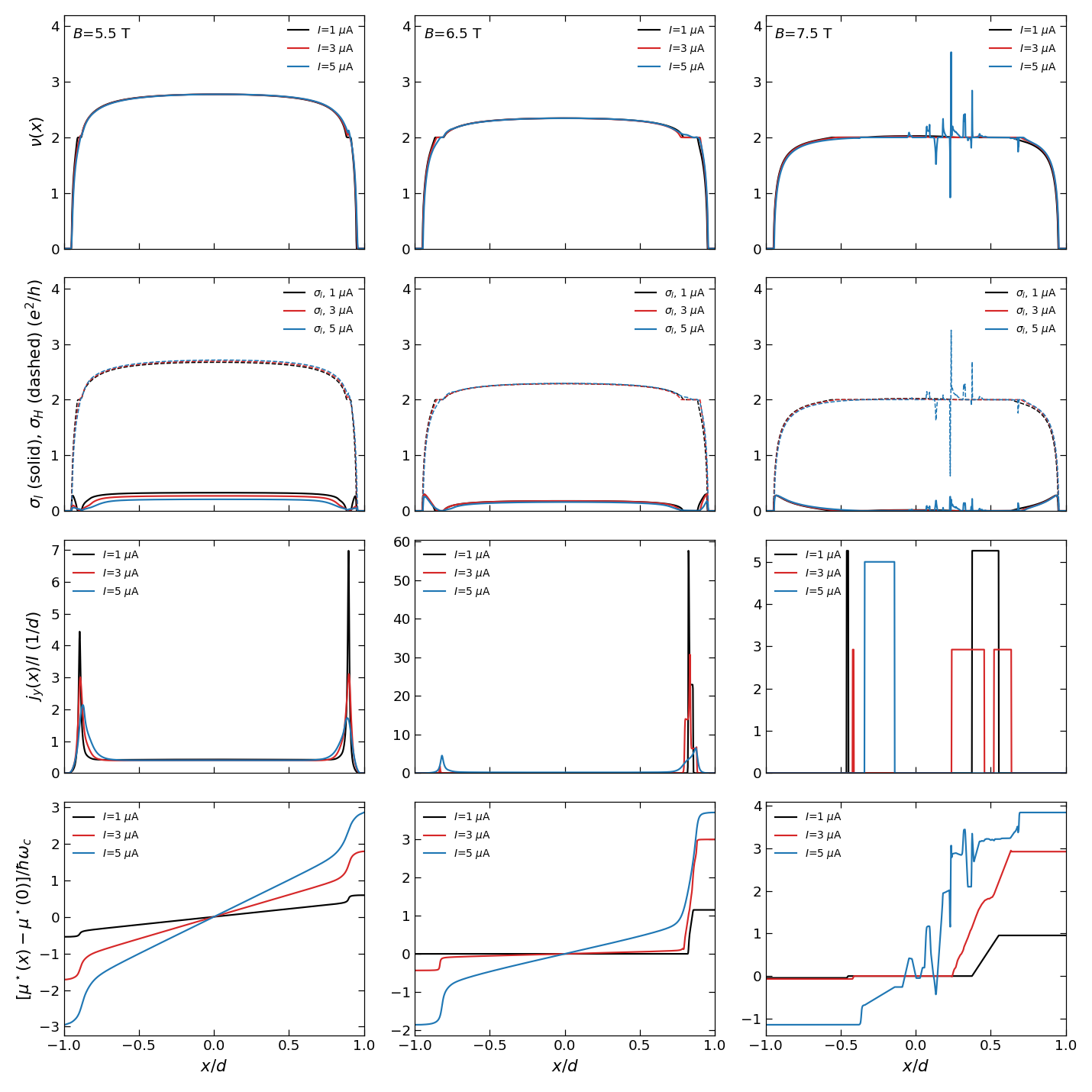}
\caption{Class B. Profiles across the bar at $B=5.5$, $6.5$ and $7.5$~T (columns) for $I=1$, $3$ and
$5\,\mu$A, with the current fed back into the electrostatics and the local energy balance.
Rows: $\nu(x)$; $\sigma_l$ (solid) and $\sigma_H$ (dashed); $j_y/I$ in units of $1/d$;
$[\mu^{*}(x)-\mu^{*}(0)]/\hbar\omega_c$. Below the plateau ($5.5$~T) the current spreads
across the compressible bulk and the edge peaks shrink with $I$. On the plateau ($6.5$~T)
the current and the Hall drop are taken by the strip at $x\simeq+0.85d$; the profile is
asymmetric under $x\to-x$ and the asymmetry grows with $I$. At the high-field edge
($7.5$~T) the incompressible bulk carries one gap at $1\,\mu$A and shows grid-scale
fine structure at $5\,\mu$A, where it is asked to sustain four (see text).}
\label{fig7}
\end{figure*}

Figure~\ref{fig7} shows $\nu(x)$, $\sigma_l$ and $\sigma_H$, $j_y/I$ and
$[\mu^{*}(x)-\mu^{*}(0)]/\hbar\omega_c$ at $B=5.5$, $6.5$ and $7.5$~T for the three currents.
The three columns are the three regimes of Fig.~\ref{fig3} read at finite current.

At $5.5$~T, below the plateau in all three runs (Fig.~\ref{fig13}), $\nu(0)\simeq2.8$, the
bulk is compressible, and the strips at $|x|\simeq0.9d$ are narrow. The current is spread
across the bar at $0.4\,I/d$ with peaks at the strips that \emph{shrink} with current ---
from $7\,I/d$ at $1\,\mu$A to $2\,I/d$ at $5\,\mu$A on the right --- and the Hall potential
is a nearly linear ramp of total height $1.15$, $3.4$ and $5.7\,\hbar\omega_c$. The strips
here are the narrow, edge-near ones of Sec.~\ref{sec:conseq}; they receive the Joule power
of Eq.~(\ref{eq:joule}) at the largest density in the bar, and their loss of weight with
increasing $I$ is the local heating acting where it was predicted to act.

At $6.5$~T, inside the $1$ and $3\,\mu$A plateaus and at the edge of the $5\,\mu$A one, the
current is concentrated in the strip at $x\simeq+0.85d$, with peaks of $57$, $31$ and
$7\,I/d$, and the Hall potential drops there in a step of $1.15$, $3.0$ and $\simeq4\,\hbar
\omega_c$; the opposite strip at $-0.85d$ carries a peak an order of magnitude smaller and a
step of $0.05$--$0.4\,\hbar\omega_c$. This is the current-induced asymmetry of
Sec.~\ref{sec:feedback}: one strip widens and takes the current and the heat, the other
narrows. At $5\,\mu$A the profile between the strips acquires a slope and the peaks broaden,
which is the leakage of current into the compressible bulk that Eq.~(\ref{eq:sigl}) predicts
once $\Te$ inside the hot strip has raised its $\sigma_l$.

At $7.5$~T, at the high-field edge, $\nu(0)$ is pinned at two and the bulk is
incompressible. At $1\,\mu$A the current sits in a flat-topped box between $x\simeq0.4d$
and $0.55d$ --- the region where the averaged $\rho_l$ is uniform --- and the potential
drops by one cyclotron gap across it. At $3\,\mu$A the drop is three gaps and the current
splits into two boxes; at $5\,\mu$A the solution develops fine structure: the $\nu=2$
segment for $0<x<0.7d$ is interrupted by narrow excursions of $\nu(x)$ and $\sigma_l$,
and the potential becomes a staircase. An incompressible region cannot sustain a potential
drop larger than the gap without changing its filling, so some subdivision under a
multi-gap Hall drop is expected on physical grounds. Whether the particular structure
shown is that subdivision or a grid-scale feature of the numerical solution is not
established here: it appears at the scale of a few cells of the $N=1001$ grid, is
sensitive to the $|\nu-2|<0.01$ threshold used for display, and lies in the regime where
$\sigma_l$ spans eight decades and the compact-support spectral functions of
Eq.~(\ref{eq:dens}) produce discontinuous derivatives. We therefore describe it as fine
structure in the numerical solution, and list in Sec.~\ref{sec:valid} the checks
(grid refinement, field step, mixing, initial state, current reversal) required before it
is given a physical reading.

\subsection{The field scan at $3\,\mu$A}\label{sec:fig8}

\begin{figure*}[t]\centering
\includegraphics[width=\textwidth]{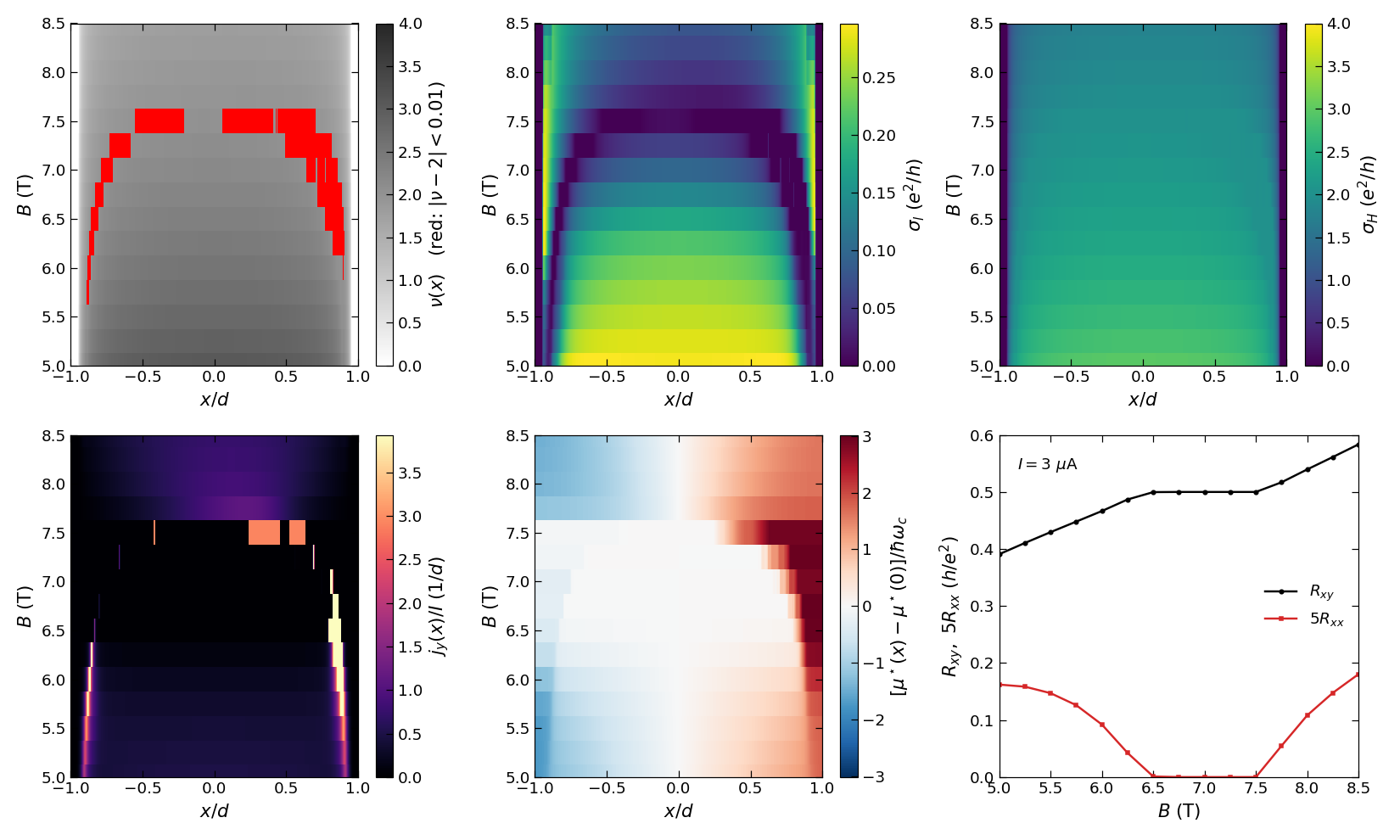}
\caption{Class B. Field scan at $I=3\,\mu$A, on linear colour scales. Top row: $\nu(x)$ with
$|\nu-2|<0.01$ in red; $\sigma_l$; $\sigma_H$. Bottom row: $j_y/I$, showing the current
migrate from both edge strips (below $6.4$~T) to the right strip alone (plateau) to the
whole bar (above $7.6$~T); $[\mu^{*}(x)-\mu^{*}(0)]/\hbar\omega_c$, the drop concentrated at
the right strip; $R_{xy}$ and $5R_{xx}$ in units of $h/e^{2}$, with the plateau from $6.5$
to $7.5$~T.}
\label{fig8}
\end{figure*}

Figure~\ref{fig8} assembles the $3\,\mu$A run as maps, on linear colour scales so that the
magnitudes can be read, together with the resistances. The $\nu=2$ region (red,
$|\nu-2|<0.01$) still forms the crescent of Fig.~\ref{fig2}, but it is no longer mirror
symmetric: on the right it is broader and ragged, on the left thin. $\sigma_l$ is low along
the crescent and across the bulk above $7.3$~T. The current map shows the point of the
section: below $6.4$~T the current flows in both edge strips, between $6.4$ and $7.5$~T it
flows almost entirely in the right strip, and above $7.6$~T it spreads across the bar. The
Hall potential map is correspondingly one-sided: the drop occurs at the right strip over the
whole plateau, reaching $3\,\hbar\omega_c$ there. The resistances read off the same run give
a plateau from $6.5$ to $7.5$~T with $R_{xy}=h/2e^{2}$ and $R_{xx}=0$, against $5.7$ to
$7.42$~T in linear response: one full tesla has been lost on the low-field side, and none on
the high-field side.

\subsection{Anatomy of the feedback at $100$~nA}\label{sec:fig9}

\begin{figure*}[t]\centering
\includegraphics[width=\textwidth]{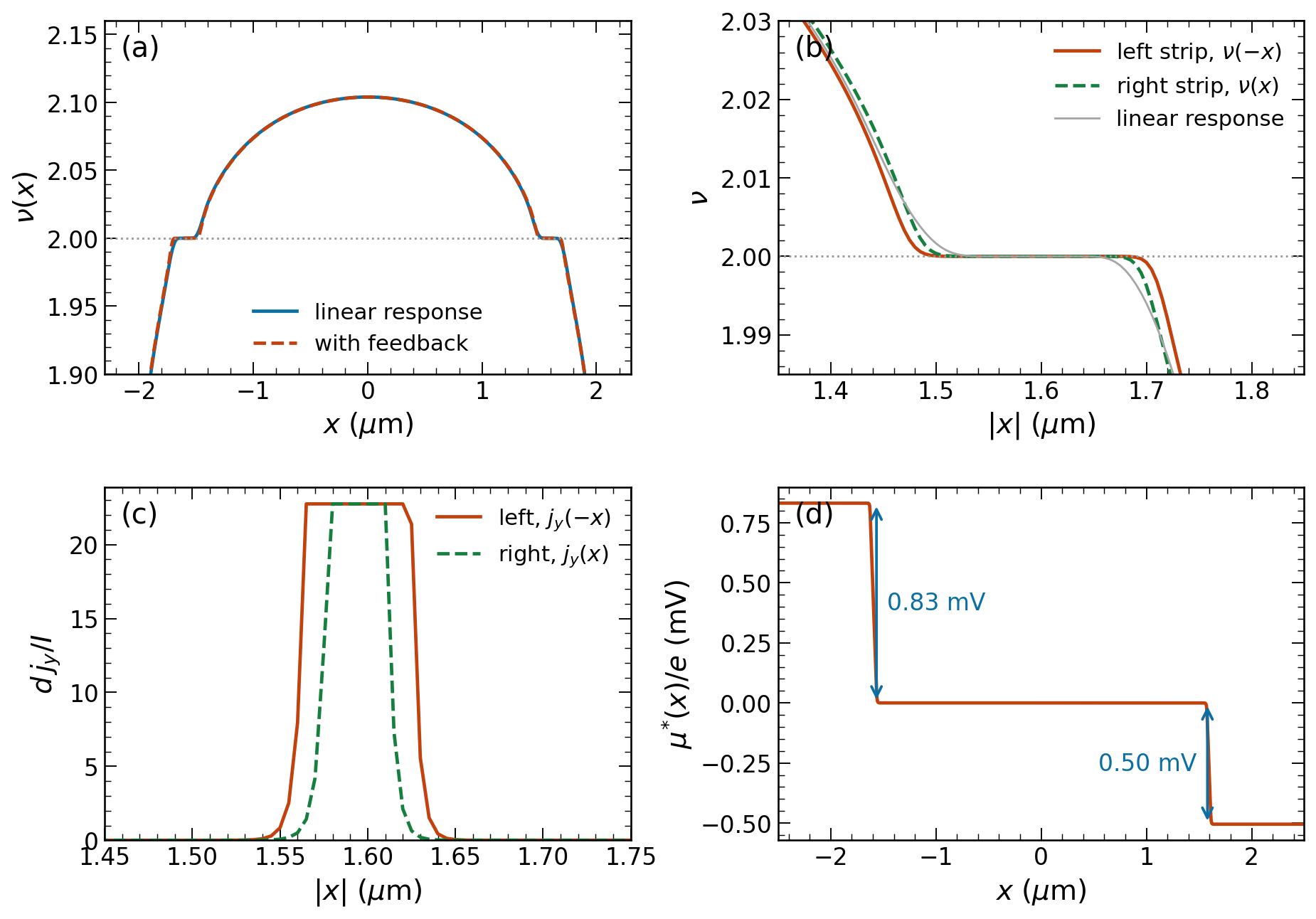}
\caption{Class B. The feedback of the current on the landscape at $I\simeq100$~nA, $\nu(0)=2.10$.
(a) $\nu(x)$ in linear response and with feedback; on this scale they coincide. (b) The
$\nu=2$ segments, with the left strip folded onto the right: the left strip is wider.
(c) $j_y/I$ in the two strips, flat-topped boxes of equal height and different width.
(d) Electrochemical potential across the bar: $0.83$~mV drops across the left strip and
$0.50$~mV across the right. The wider strip carries the larger current and receives the
larger share of the Joule power.}
\label{fig9}
\end{figure*}

Before the current series is pressed further, Fig.~\ref{fig9} isolates the mechanism at a
current small enough that the two strips can be resolved separately: $I\simeq100$~nA
($V_H=1.33$~mV, $eV_H/\hbar\omega_c=0.11$) at $\nu(0)=2.10$. This run uses a wider bar
than the $d=1.5\,\mu$m geometry of the rest of the paper, so that the strips are well
separated from the depleted edges; it is an illustration of the mechanism, not a control
calculation for Figs.~\ref{fig7}--\ref{fig13}, and the numbers below are specific to it.
Panel~(a) shows that on the scale of the whole profile the feedback changes nothing:
linear response and the self-consistent driven state coincide. Panel~(b) zooms into the
$\nu=2$ segments and folds the left strip onto the right by $x\to-x$: the left strip,
$1.49<|x|<1.73\,\mu$m, is wider than the right, $1.50<|x|<1.70\,\mu$m, and the
linear-response strip lies between them. Panel~(c) shows the current in the two strips as
flat-topped boxes of equal height $22.7\,I/d$ --- the height is set by the uniform
$\rho_l$ floor inside a strip --- whose widths differ, so that the left strip carries
$62\%$ of the current. Panel~(d) is the consequence for the Hall potential: $0.83$~mV drops
across the left strip and $0.50$~mV across the right. The strip at the higher
electrochemical potential is the wider one, carries the larger current and, by
Eq.~(\ref{eq:joule}), receives $62\%$ of the Joule power. Reversing $I$ interchanges the two
exactly. This is the current-induced asymmetry predicted by the screening theory
\cite{Siddiki2009EPL87}, computed by Gerhardts, Panos and Weis \cite{Gerhardts2013} and
observed in gate-defined bars \cite{Siddiki2009EPL88,Siddiki2010NJP}; the addition here is
that it fixes the side of the bar that heats.

\subsection{The current series: maps at $1$, $3$ and $5\,\mu$A}\label{sec:maps}

\begin{figure*}[t]\centering
\includegraphics[width=\textwidth]{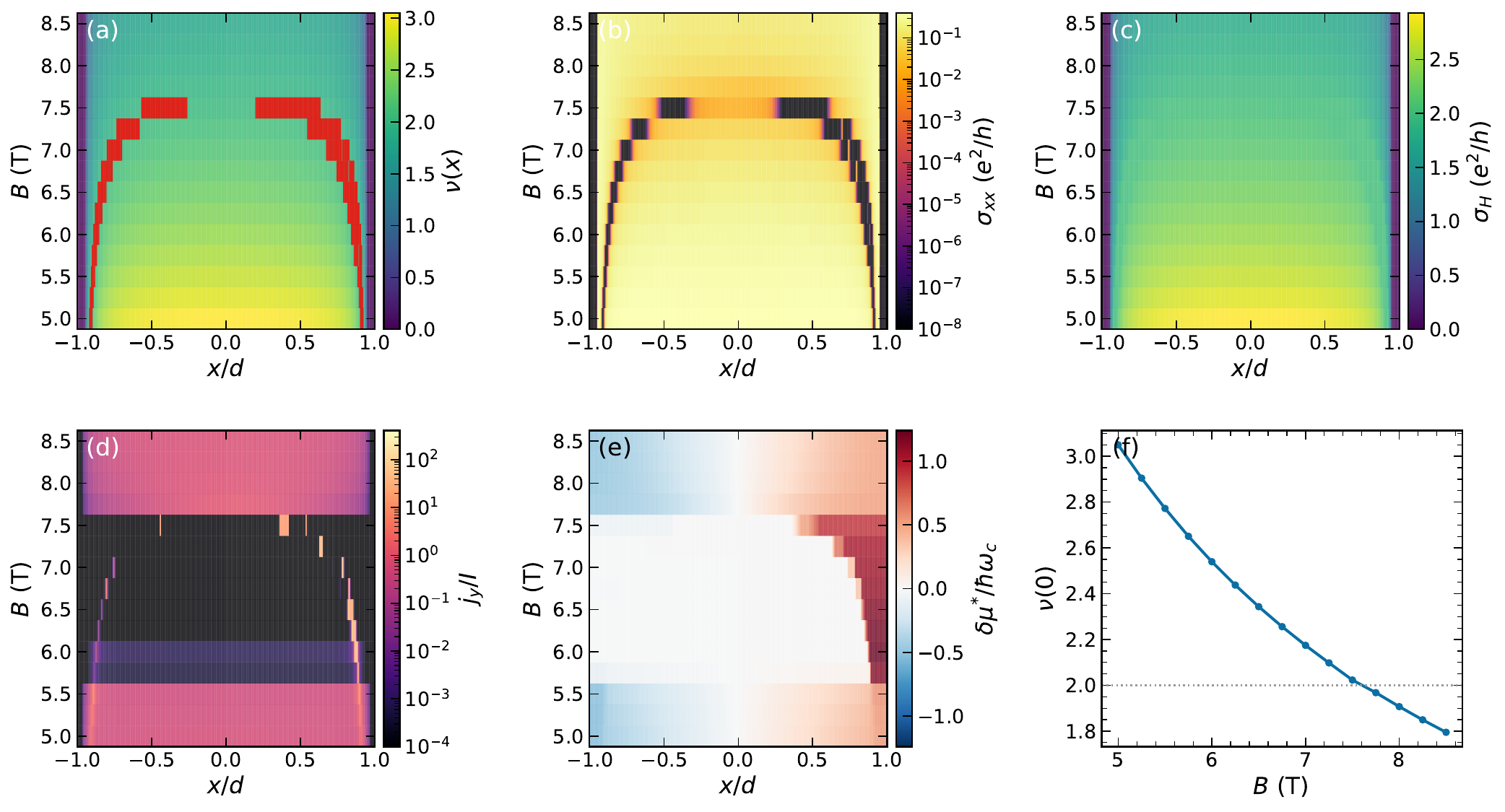}
\caption{Class B. Field scan at $I=1\,\mu$A ($eV_H/\hbar\omega_c\simeq1$) on logarithmic colour
scales. (a) $\nu(x)$ with the $\nu=2$ region in red; (b) $\sigma_{xx}$; (c) $\sigma_H$;
(d) $j_y/I$; (e) $\delta\mu^{*}/\hbar\omega_c$; (f) $\nu(0)$ versus $B$. The current
leaves the bulk for the strips at $5.6$~T and returns to it above $7.6$~T.}
\label{fig10}
\end{figure*}

\begin{figure*}[t]\centering
\includegraphics[width=\textwidth]{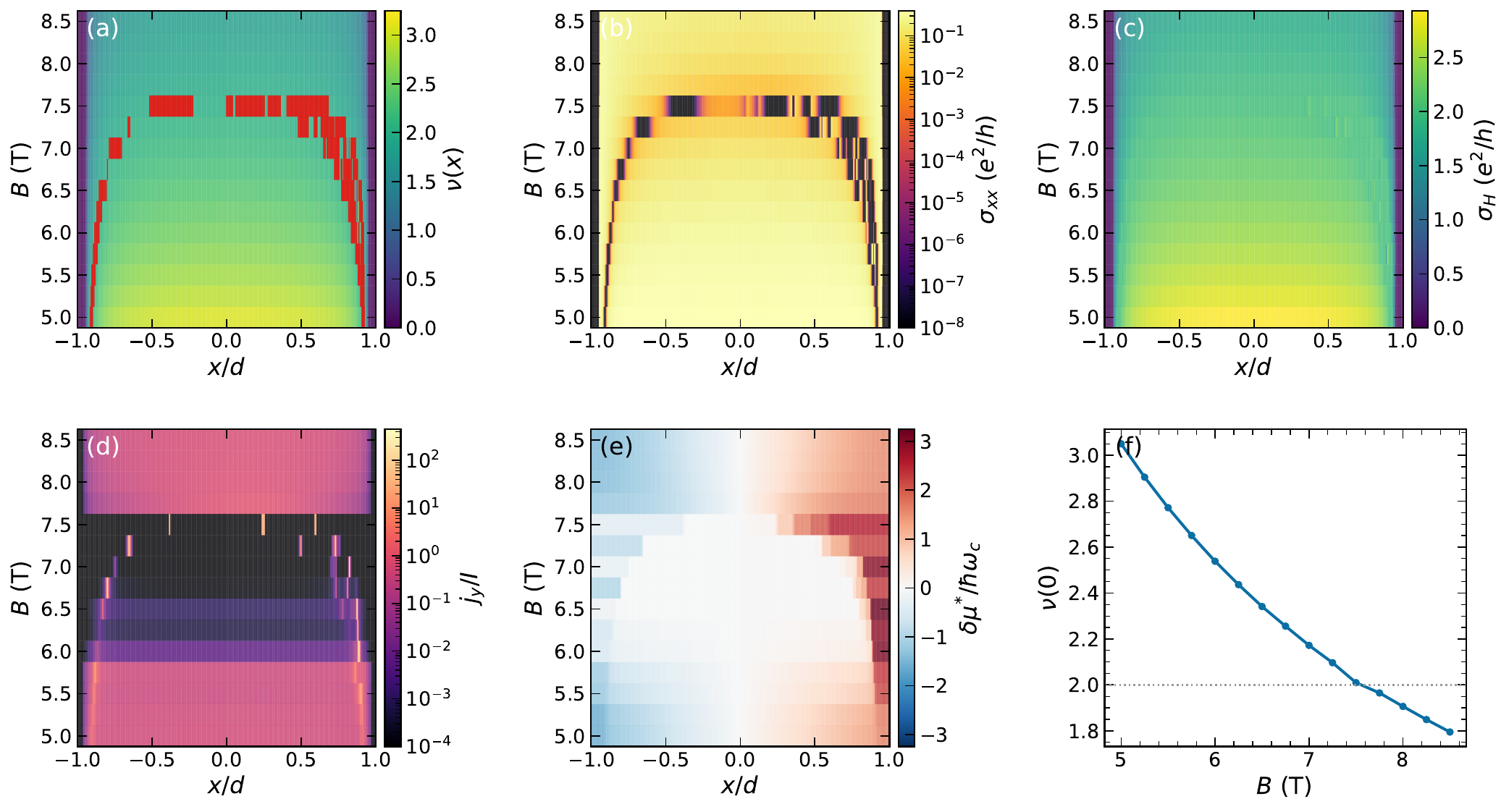}
\caption{Class B. As Fig.~\ref{fig10}, at $I=3\,\mu$A ($eV_H/\hbar\omega_c\simeq3$). The right arm
of the $\nu=2$ crescent thickens and roughens, $\sigma_{xx}$ inside it is raised by the
local heating, the current leaves the bulk only at $5.9$~T, and the central filling factor is
pinned at two near $7.5$~T.}
\label{fig11}
\end{figure*}

\begin{figure*}[t]\centering
\includegraphics[width=\textwidth]{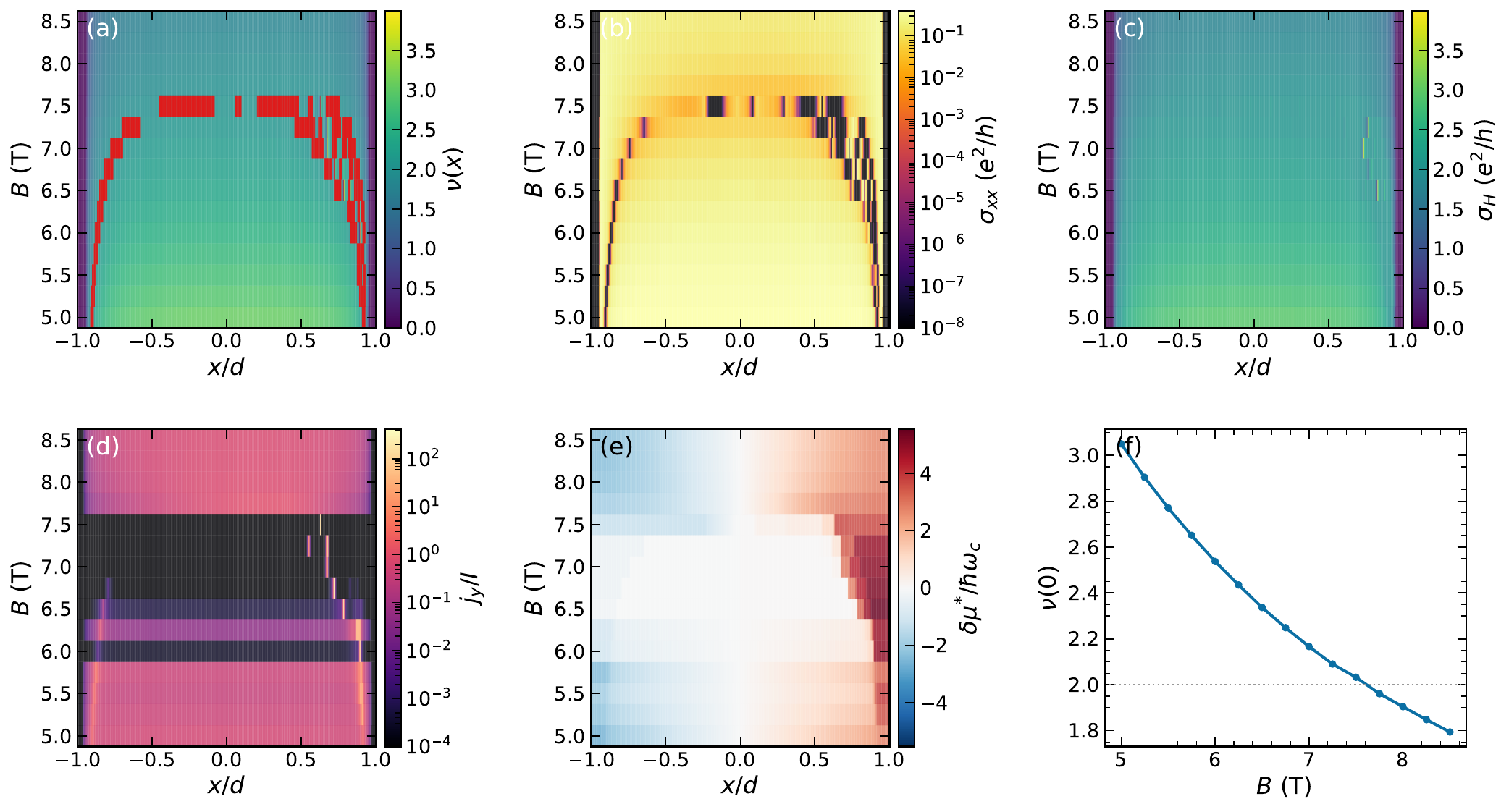}
\caption{Class B. As Fig.~\ref{fig10}, at $I=5\,\mu$A ($eV_H/\hbar\omega_c\simeq5$). The right arm
of the incompressible region shows fine structure from $6.4$~T upwards and the bulk region
at $7.5$~T is perforated (binary colouring, see text); the current enters the strips at $5.9$~T, returns to the bulk
between $6.1$ and $6.5$~T as the hot strips lose their weight, and settles into the strips
only above $6.5$~T.}
\label{fig12}
\end{figure*}

Figures~\ref{fig10}, \ref{fig11} and~\ref{fig12} are the field scans at $1$, $3$ and
$5\,\mu$A on logarithmic colour scales, with the central filling factor in panel~(f). They
share the reference state $\nu(0)B=15.2$~T, so $\nu(0)$ passes through two at $7.6$~T in
all three. Read as a sequence they show the heating loop act.

\emph{The incompressible region.} At $1\,\mu$A (Fig.~\ref{fig10}a) the red crescent is nearly
mirror symmetric and closes cleanly across the bulk at $7.5$~T. At $3\,\mu$A
(Fig.~\ref{fig11}a) the right arm thickens and roughens above $6.5$~T while the left thins;
at $5\,\mu$A (Fig.~\ref{fig12}a) the right arm dissolves into scattered $\nu=2$ pixels
from $6.4$~T upwards and the closed bulk region at $7.5$~T is perforated. The binary
colouring exaggerates grid-scale structure, and the same caveat as in
Sec.~\ref{sec:fig7} applies.

\emph{The conductivity.} Panel~(b) shows $\sigma_{xx}$ falling to $10^{-8}\,e^{2}/h$ along
the crescent at $1\,\mu$A. At $3$ and $5\,\mu$A the dark arc is interrupted: along the right
arm $\sigma_{xx}$ inside the strip is orders of magnitude above its $1\,\mu$A value. This is
Eq.~(\ref{eq:sigl}) with $\Te>\TL$ in the strip that carries the current --- the activation of
$\sigma_l$ by the local electron temperature --- and it is the quantity that the Akera
picture identifies as the controlling one \cite{Akera2000,Akera2002}.

\emph{The current.} Panel~(d) records where the current is, and the low-field boundary of the
plateau can be read off it. At $1\,\mu$A the current leaves the bulk for the strips at
$5.6$~T; at $3\,\mu$A at $5.9$~T; at $5\,\mu$A it enters the strips at $5.9$~T, returns to
the bulk between $6.1$ and $6.5$~T, and only then settles into the strips. That return is
the signature of the bootstrap of Sec.~\ref{sec:heat}: the strips at $6.1$--$6.5$~T are
narrow, they take the current and the power, their $\sigma_l$ rises, and they hand the
current back to the compressible bulk. Above $7.6$~T, where no strip exists, the current is
spread across the bar in all three runs and the maps coincide.

\emph{The Hall potential.} Panel~(e), in units of $\hbar\omega_c$, shows the drop
concentrated on the right strip over the plateau at all three currents and growing from one
to five gaps. At $3$ and $5\,\mu$A the drop at $7.5$~T extends inwards from the right edge
to $x\simeq0.3d$, the staircase of Fig.~\ref{fig7}.

\emph{The central filling factor.} Panel~(f) is the same curve in all three, except for a
short flat segment at $\nu(0)=2$ near $7.5$--$7.6$~T, most visible at $3\,\mu$A: the
central density is briefly pinned at integer filling by the Hall potential, the
one-dimensional analogue of the pinning that produces the plateau itself.

\subsection{Resistances versus current: a one-sided loss of plateau}\label{sec:fig13}

\begin{figure}[t]\centering
\includegraphics[width=\columnwidth]{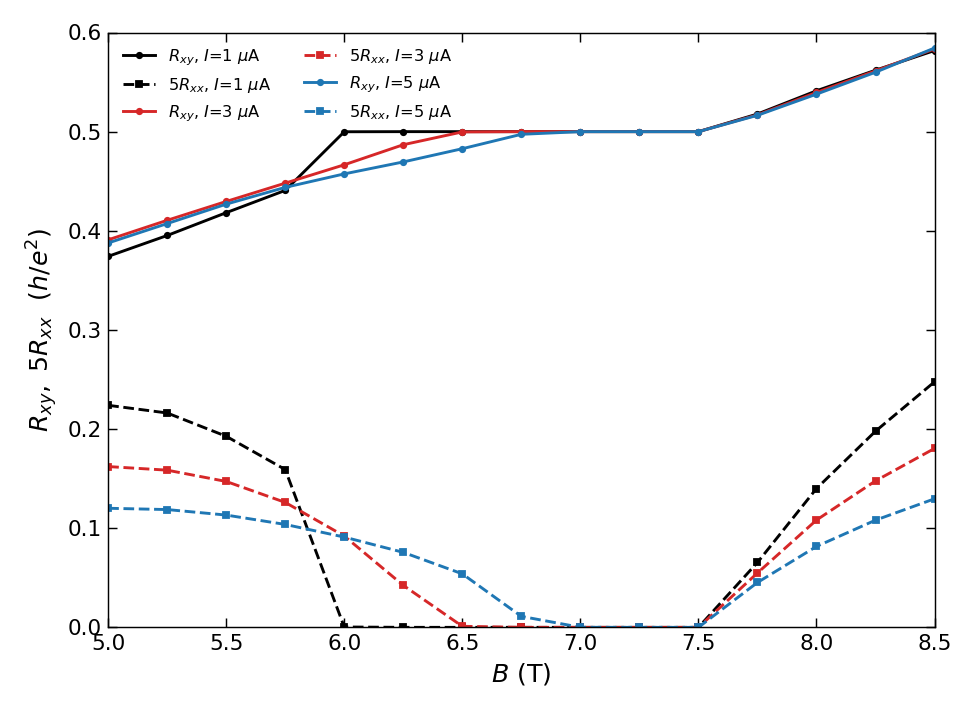}
\caption{Class B. $R_{xy}$ (solid) and $5R_{xx}$ (dashed) in units of $h/e^{2}$ versus field for
$I=1$, $3$ and $5\,\mu$A. The quantised value and the high-field edge, at $\nu(0)=2$, are
unchanged by current; the low-field edge moves from $6.0$ to $6.5$ to $7.0$~T. Below the
plateau $R_{xx}$ decreases with current (hotter compressible bulk); on the plateau it
increases with current (hotter strips).}
\label{fig13}
\end{figure}

Figure~\ref{fig13} collects $R_{xy}$ and $5R_{xx}$ for the three currents. The quantised
value is unchanged: on every plateau $R_{xy}=h/2e^{2}$ to the resolution of the plot. The
high-field edge is unchanged: all three curves leave the plateau between $7.5$ and $7.75$~T,
where $\nu(0)$ crosses two. The low-field edge moves monotonically with current: $R_{xx}$
reaches zero at $6.0$~T for $1\,\mu$A, $6.5$~T for $3\,\mu$A and $7.0$~T for $5\,\mu$A, and
$R_{xy}$ reaches $h/2e^{2}$ at the same fields. Expressed through $\nu(0)$ the plateau
occupies $2.0\lesssim\nu(0)\lesssim2.55$, $2.35$ and $2.15$ respectively, against $2.6$ in
linear response and $2.45$ at $408$~nA (Fig.~\ref{fig6}). The plateau is eaten from its
low-field end at a rate of roughly $0.25$~T per microampere, and not at all from the other.

Below the plateau $R_{xx}$ itself falls with current --- $0.045$, $0.032$ and
$0.024\,h/e^{2}$ at $5$~T --- because the bulk is compressible there, the heat is spread
across the bar and a hotter compressible bulk conducts better; above the plateau the order
is the same. On the plateau the ordering is reversed: at any field between $6$ and $7$~T
the higher current has the larger $R_{xx}$, because there the heat is deposited in a strip
of width $a_k$ rather than across $2d$, and a hotter strip conducts worse in the sense that
matters, namely $\rho_l$ rises and the strip loses its weight.

\emph{An operational breakdown criterion.} Several phenomena have been called
breakdown above --- $R_{xx}$ becoming finite, the current leaving a strip, the loss of the
$R_{xy}$ plateau, fine structure in the $\nu=2$ region, $\varepsilon_{\rm ad}\to1$, a
rise of $\Te$, $v_D\to v_s$ --- and they are related but not identical. We adopt one:
at a given field, the critical current $I_c(B)$ is the smallest current at which
$R_{xx}$ exceeds $10^{-3}\,h/e^{2}$ ($26\,\Omega$, the resolution of Fig.~\ref{fig13});
equivalently, at given $I$, the low-field plateau edge $B_c(I)$ is where $R_{xx}$ crosses
that value. On this definition $B_c=6.0$, $6.5$ and $7.0$~T at $1$, $3$ and $5\,\mu$A, and
inverting, $I_c\simeq1\,\mu$A at $6.0$~T and $\simeq3\,\mu$A at $6.5$~T, i.e.\
$j_c=I_c/2d\simeq0.3$--$1$~A/m --- inside the experimental range of
Table~\ref{tab:scales}, though the comparison is only as good as $\tau_0$. The other
signatures are then to be compared with this one: on the maps the current leaves the
strips (Figs.~\ref{fig10}--\ref{fig12}d) within one field step of $B_c$, which is the
consistency between the local and the integrated description that Eq.~(\ref{eq:RHRL})
guarantees; the relation of $B_c$ to the maximum of $\Te$ is part of Sec.~\ref{sec:valid}.

\emph{Alternative mechanisms.} The electron-heating (bootstrap) picture is not the only
account of breakdown. In samples with long-range disorder the incompressible region is
threaded by compressible puddles, and breakdown has been described as a field-driven
percolation of these puddles --- the mechanism argued for in the quantum-anomalous-Hall
films of Ref.~\cite{Lippertz2022} --- or as inter-Landau-level tunnelling where the local
field is steepest (QUILLS) \cite{Eaves1986,Nachtwei1999}. The question is open: nanoscale
magnetic imaging of the same class of films has since attributed the breakdown to
electron heating, with the electrons driven out of equilibrium at the hot spots and then
throughout the device \cite{PNAS2026}. The present model, with disorder only
through collision broadening, cannot represent the puddle route, and the results above do
not exclude it. What the two routes predict differently, and what can be computed in an
extended version of the present model, is: (i) the scaling of $I_c$ with bar width ---
linear in the thermal picture, since the power per unit strip area is what matters, and
set by the puddle geometry rather than by $2d$ in the percolative one; (ii) hysteresis and
a time-dependent onset on the scale of $\tau_\epsilon$ (thermal) versus a
history-independent, essentially instantaneous onset (percolative); (iii) a lattice-
temperature dependence of $I_c$ through $P_{\rm loss}(\Te,\TL)$ (thermal) that is absent
or weak in the field-driven picture; and (iv) the location of the first dissipation ---
inside the strip that carries the current (thermal) or at the narrowest bottleneck between
puddles (percolative), which is a question that scanning thermometry can answer. On
present evidence the calculation shows a current-driven, one-sided plateau narrowing; it
does not adjudicate between the two mechanisms.

\emph{Comparison with current and dissipation imaging.} The predictions of this section
concern quantities that are now imaged directly. NV-centre magnetometry maps current density in two-dimensional conductors at
sub-micrometre resolution \cite{Tetienne2017,Chang2017}, SQUID-on-tip magnetometry
resolves current distributions at $\sim50$~nm \cite{Uri2020}, and SQUID-on-tip
thermometry resolves dissipation at comparable scales \cite{Halbertal2016,Marguerite2019}.
The averaging length $\lambda=30$~nm of Sec.~\ref{sec:avg} is below the resolution of all
of these, so the strip width $a_k$, not $\lambda$, is the quantity imaging tests. Against such data the model makes four quantitative
statements: the width of the current-carrying strip ($249$~nm in linear response,
widening under current, Fig.~\ref{fig9}c); the fraction of the total current in the strips
($>99.99\%$ on the plateau, falling at $B_c$); the migration of the strip with $B$
(Fig.~\ref{fig2}d), already seen by scanning-force microscopy in GaAs
\cite{Ahlswede1,Ahlswede2}; and, under current, the concentration of both current and
Hall drop in one strip (Figs.~\ref{fig8}, \ref{fig9}) with dissipation, by
Eq.~(\ref{eq:joule}), in the same place. The graphene measurements of
Ref.~\cite{Marguerite2019} found dissipation at resonant edge scatterers rather than by
local Joule heating, in a geometry with sharp, reconstructed edges; the smooth-edge GaAs
case treated here has not been imaged at that resolution, and the location of the
dissipation relative to the strip is the cleanest discriminating measurement.

Taken together, Figs.~\ref{fig6}--\ref{fig13} are consistent with the expectation of
Sec.~\ref{sec:conseq}. Current acts on the plateau through the strips that are narrow,
those are the strips near the edges at low field, and the plateau is lost on that flank;
for the GaAs parameters, bar width and edge confinement used here the high-field edge did
not move within the resolution of the scan --- an observation for this parameter set, not a
general property; its dependence on width, confinement, disorder and injection geometry
has not been scanned. The proposed mechanism is that the current
chooses the strip, the strip receives the Joule power at a density set by the same
$1/\rho_l$ weight, its suppressed $\kappa_{xx}$ converts the power into electron
temperature, and the temperature raises the strip's $\sigma_l$ until the current leaves it.
Two ingredients of that chain --- the redistribution of current and Hall drop between the
strips, and the current dependence of $B_c$ --- are displayed in the figures; the third,
the electron-temperature field itself, is not shown in Figs.~\ref{fig7}--\ref{fig13} and
must be, together with the checks below, before the narrowing is attributed to heating
rather than to the electrostatic feedback alone, which also acts on narrow strips.

\subsection{Validation of the thermal solution}\label{sec:valid}

\begin{figure*}[t]\centering
\includegraphics[width=\textwidth]{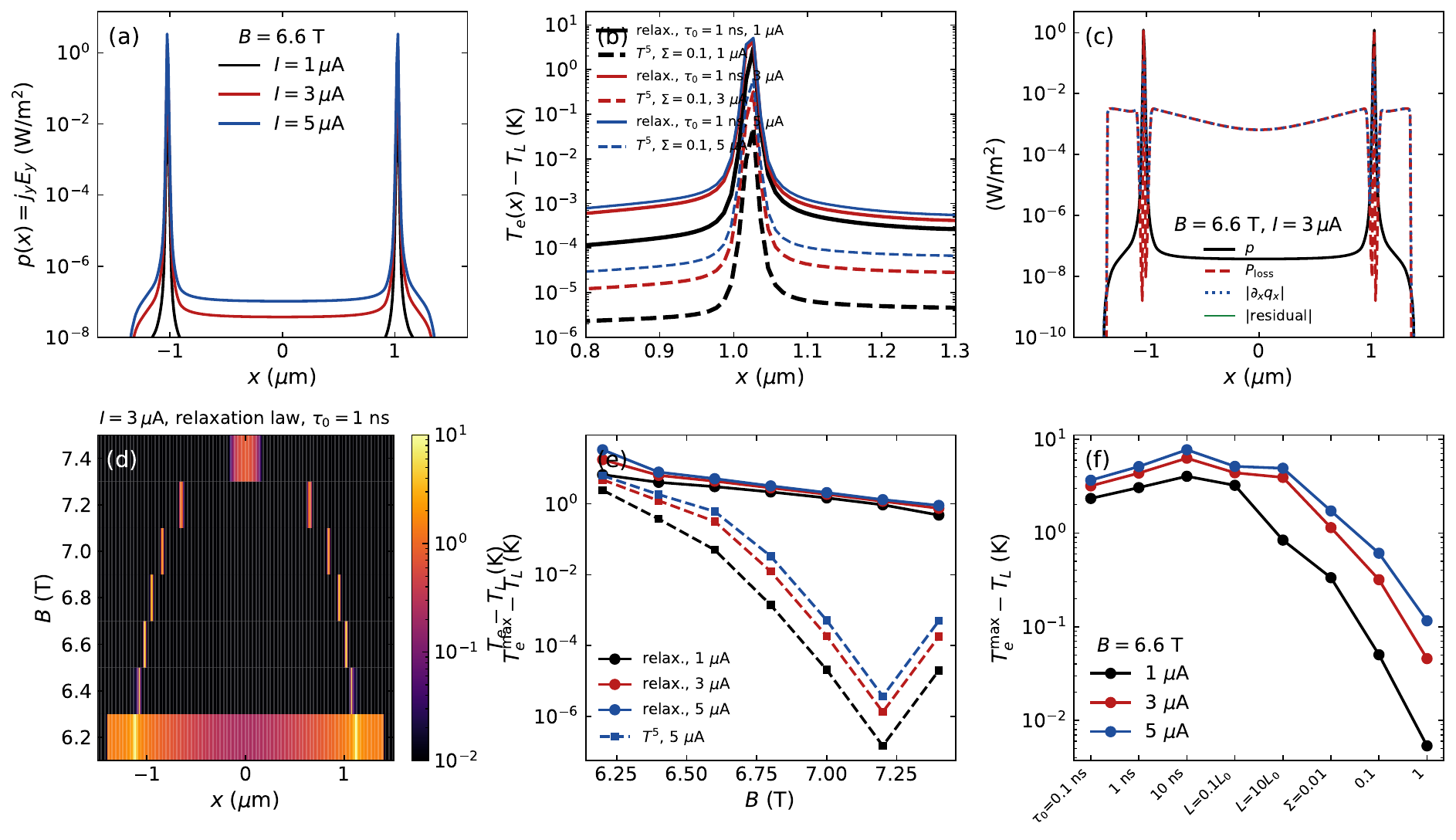}
\caption{Class-C thermal solution, post-processed on the equilibrium profile
($d=1.5\,\mu$m, $N=301$, $\TL=1.5$~K). (a) Joule power density $p(x)=j_yE_y$ at $6.6$~T for $I=1$, $3$, $5\,\mu$A.
(b) $\Te(x)$ from Eq.~(\ref{eq:balance}), linearised relaxation law with $\tau_0=1$~ns (solid; the internal-energy form (\ref{eq:lossU}) raises $\Te^{\max}$ by $10$--$30\%$, Table~\ref{tab:budget}) and
$T^{5}$ law with $\Sigma=0.1$~W\,m$^{-2}$K$^{-5}$ (dashed, $5\,\mu$A). (c) Energy balance
at $3\,\mu$A: $p$, $P_{\rm loss}$, $|\partial_xq_x|$ and the pointwise residual.
(d) $\Te(x,B)$ at $3\,\mu$A, relaxation law. (e) $\Te^{\max}-\TL$ along the scan for the
three currents and both closures. (f) Sensitivity of $\Te^{\max}-\TL$ at $6.6$~T to
$\tau_0$, to the Lorenz number and to $\Sigma$. Grid dependence: Table~\ref{tab:grid}.}
\label{fig14}
\end{figure*}

\begin{figure*}[t]\centering
\includegraphics[width=\textwidth]{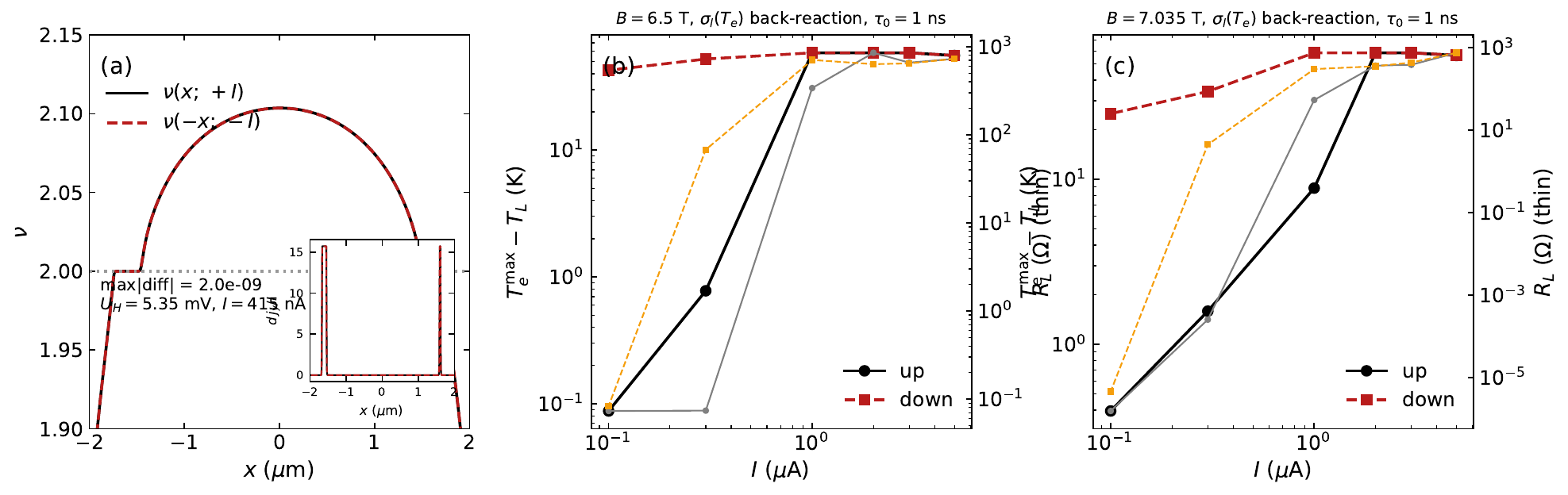}
\caption{(a) Class B: current-reversal test on the self-consistent driven state of the
$d=2.5\,\mu$m bar at $\nu(0)=2.10$, $U_H=5.35$~mV: $\nu(x;+I)$ against $\nu(-x;-I)$
(inset: $j_y$). (b), (c) Class C$'$: current sweeps at $6.5$ and $7.035$~T with the
$\sigma_l(\Te)$ back-reaction and the linearised relaxation closure, upward from the cold state
(circles) and downward from the hot state (squares); thick lines $\Te^{\max}-\TL$, thin
lines $R_L$. The cold branch is lost between $0.3$ and $1\,\mu$A at $6.5$~T and between $1$ and $2\,\mu$A at $7.035$~T; points at the solver cap
are outside the validity of the frozen profile and are shown only to mark the loss of the cold branch.}
\label{fig15}
\end{figure*}

Figures~\ref{fig14} and~\ref{fig15} report the checks that the thermal description
requires. They are computed with the model of Secs.~\ref{sec:heat}--\ref{sec:num} on the
equilibrium density profile, so they test the energy balance and its sensitivity, not the
full feedback loop of Figs.~\ref{fig7}--\ref{fig13}.

\emph{Where the heat is and where it goes.} Figure~\ref{fig14}(a) shows $p(x)=j_yE_y$ at
$6.6$~T for $1$, $3$ and $5\,\mu$A: the power is deposited in the two strips at
$|x|\simeq1.03\,\mu$m, seven decades above the compressible bulk, and scales as $I^{2}$.
Panel~(b) shows the resulting $\Te(x)$. With the relaxation law the strips sit at
$4.6$--$6.6$~K on a $1.5$~K background; with the $T^{5}$ law they rise by
$0.05$--$0.6$~K. In both cases the hot region is the strip that carries the current, and
the compressible bulk stays at $\TL$: the electronic thermal conductivity
$\kappa_{xx}=L\Te\sigma_l$ is suppressed inside the strip by the same factor as
$\sigma_l$, so the heat cannot leave it sideways. Panel~(c) shows the balance itself for
$3\,\mu$A: $P_{\rm loss}$ follows $p$ across the strip, the conduction term
$|\partial_xq_x|$ is three to five decades smaller except at the strip edges, and the
pointwise residual is below $10^{-10}$ of the peak power. Panel~(d) is the map
$\Te(x,B)$ at $3\,\mu$A: the hot arcs are the current-carrying arcs of
Fig.~\ref{fig2}(d), and they warm towards low field, where the strips narrow and $R_L$
rises. Panel~(e) collects $\Te^{\max}-\TL$ along the scan: it rises from $10^{-4}$~K
($T^{5}$) or $1$~K (relaxation) at the plateau centre to $1$--$2$~K ($T^{5}$) or
$4$--$8$~K (relaxation) at $6.4$~T, so that in either closure the heating is largest on
the low-field flank, as Sec.~\ref{sec:conseq} anticipated. Below $6.2$~T the strips are
narrower than the $N=301$ grid resolves ($10$~nm cells, $\lambda=30$~nm) and the scan is
not continued; the $N=1001$ grid of Figs.~\ref{fig10}--\ref{fig13} is needed there.

\emph{Sensitivity.} Panel~(f) varies the closure at $6.6$~T. For the relaxation law,
$\tau_0=0.1$, $1$ and $10$~ns give $\Te^{\max}-\TL=2.3$, $3.1$ and $4.0$~K at $1\,\mu$A
and $3.7$, $5.1$ and $7.7$~K at $5\,\mu$A --- a weak, roughly logarithmic dependence,
because the activated $c_{\rm e}$ rises steeply with $\Te$ and absorbs most of a change
in $\tau_0$. The Lorenz number matters little ($L=0.1L_0$ changes $\Te^{\max}$ by
$6\%$; $L=10L_0$ lowers it at $1\,\mu$A by allowing conduction to the strip edge, and
not at $5\,\mu$A). The $T^{5}$ law with $\Sigma=0.01$, $0.1$ and $1$~W\,m$^{-2}$K$^{-5}$
gives $0.33$, $0.05$ and $0.005$~K at $1\,\mu$A, the expected $\Sigma^{-1/5}$ scaling.
The largest uncertainty is therefore not a parameter but the form of the loss law
inside a strip: a factor of $10$--$60$ in $\Te^{\max}-\TL$ separates the two closures at
the same power, and a microscopic $P_{\rm loss}$ for a gapped, broadened Landau level
\cite{AkeraSuzuura2005,Kanamaru2006} is what would remove it. \emph{Grid convergence.} Table~\ref{tab:grid} repeats the $1\,\mu$A thermal solution at
$6.4$ and $6.6$~T on grids of $N=301$, $601$ and $1001$ cells. Between $N=601$ and $1001$ the strip width and $\Te^{\max}$ change by $2\%$ or less at both fields, while $R_L$, which is set by the exponentially small $\sigma_l$ inside a strip of only ten to twelve cells, still changes by a factor $1.4$; $R_H$ is converged to $10^{-5}$. The earlier $N=451$ outlier quoted in a previous version of this work ($R_L=0.84\,\Omega$) was a strip straddling a cell boundary and is superseded. The temperatures are therefore converged at the few-per-cent level from $N=601$ upwards; the resistance on the flank, and hence $B_c(I)$ read from an $R_{xx}$ threshold, is not yet converged to better than a factor of two. A residual of
$10^{-10}$ shows only that the discretised equation was solved; the table is what shows
how far the discretisation represents the continuum problem, and the flank values of
Table~\ref{tab:budget} and Fig.~\ref{fig14} are to be read with its spread.

\begin{table*}[t]
\caption{Grid dependence of the class-C thermal solution at $I=1\,\mu$A (internal-energy
relaxation law, $\tau_0=1$~ns).}

\begin{ruledtabular}\begin{tabular}{llllllll}
$B$ (T) & $N$ & strip width (nm) & $R_L$ ($\Omega$) & $R_H$ ($h/e^{2}$) & $p_{\max}$ (W/m$^{2}$) & $\Te^{\max}$ (K) & \\
\hline
$6.4$ & $301$ & $100$ & $9.9\times10^{-2}$ & $0.49956$ & $1.6$ & $6.62$ \\
$6.4$ & $601$ & $100$ & $7.7\times10^{-2}$ & $0.49956$ & $2.0$ & $7.58$ \\
$6.4$ & $1001$ & $102$ & $5.6\times10^{-2}$ & $0.49955$ & $1.7$ & $7.48$ \\
$6.6$ & $301$ & $110$ & $9.7\times10^{-3}$ & $0.50010$ & $0.14$ & $5.39$ \\
$6.6$ & $601$ & $120$ & $3.1\times10^{-3}$ & $0.50010$ & $0.077$ & $5.47$ \\
$6.6$ & $1001$ & $123$ & $2.2\times10^{-3}$ & $0.50010$ & $0.061$ & $5.37$ \\
\end{tabular}\end{ruledtabular}
\label{tab:grid}
\end{table*}

\emph{Adiabaticity under the Hall field.} The adiabatic parameter with the driven potential,
$\varepsilon_{\rm ad}=\ell|\partial_x(V-\mu^{*})|/\hbar\omega_c$, is the quantity that
decides whether the local description is controlled at large current. On the
self-consistent driven state of the $d=2.5\,\mu$m bar at $U_H=0.42\,\hbar\omega_c$
(class~B, Fig.~\ref{fig15}a) its maximum inside the strips is $0.051$, against $0.036$
in equilibrium, and $0.21$ at the compressible edge; the feedback widens the strip
that takes the drop and keeps the gradient small. On the frozen class-A profile the same
quantity is $0.29$--$0.37$ at $1\,\mu$A, $0.78$--$0.96$ at $3\,\mu$A and $1.3$--$1.5$ at
$5\,\mu$A across the plateau: without the widening, a Hall drop of three gaps across a
$200$~nm strip already brings $\varepsilon_{\rm ad}$ to unity. The class-B runs of
Figs.~\ref{fig7}--\ref{fig13} do widen the strips, but their $\varepsilon_{\rm ad}(x)$
has not been extracted, and until it is the $5\,\mu$A results must be regarded as
outside the controlled regime of Sec.~\ref{sec:bo}. Landau-level mixing at
$\varepsilon_{\rm ad}\sim1$ is, of course, the non-adiabatic channel of
Sec.~\ref{sec:breakdown}.

\emph{Current reversal.} Figure~\ref{fig15}(a) is the test of Sec.~\ref{sec:feedback} on
the self-consistent driven state at $U_H=5.35$~mV ($I=415$~nA, the drive of
Fig.~\ref{fig6}): $\nu(x;I)$ and $\nu(-x;-I)$ coincide to $2\times10^{-9}$ in $\nu$, and
the current profiles to $3\times10^{-8}$ of the peak. Under $I\to-I$ the wide strip
($293$~nm, $80\%$ of the current, $4.3$~mV of the Hall drop) and the narrow strip
($213$~nm, $20\%$, $1.0$~mV) interchange exactly. The asymmetry of Figs.~\ref{fig7}--
\ref{fig9} is therefore the driven boundary condition, not the continuation.

\emph{Stability of the cold branch.} Panels~(b) and~(c) close the loop
$\Te\to\sigma_l(\Te)\to E_y^{0}\to p\to\Te$ at fixed current and sweep $I$ upward from
the cold state and downward from the hot one, with the relaxation closure. At $6.5$~T the
cold branch exists up to $0.3\,\mu$A ($\Te^{\max}=2.3$~K, $R_L=0.07\,\Omega$) and is lost
before $1\,\mu$A, where no stationary solution with $k_B\Te<\Gamma$ ($\Te<17$~K, the
point at which thermal smearing of the strip density, held frozen here, would exceed the
level broadening) exists any longer; the solver then runs to its cap and $R_L$ rises to
hundreds of ohms, but that state is outside the model and is not reported as a result. At the plateau centre, $7.035$~T, the cold branch survives to $1\,\mu$A, but by then
the bootstrap is visibly under way --- $\Te^{\max}=10.4$~K and $R_L$ raised from
$0.3$~m$\Omega$ at $0.3\,\mu$A to $54\,\Omega$, with $R_H$ beginning to leave
$h/2e^{2}$ --- and it is lost before $2\,\mu$A. The downward sweep stays on the hot
branch: the transition is hysteretic, which is the bistability of
Refs.~\cite{Komiyama2000,Akera2000}. Two readings must be kept apart. The
loss of the cold branch at $0.3$--$1\,\mu$A on the flank and $1$--$2\,\mu$A at the centre
is a model-dependent estimate: it holds for the relaxation closure with activated heat
capacity (the same in its linearised and internal-energy forms, the latter shifting the
thresholds down by at most one step of the sweep), on a frozen profile, without contacts
or longitudinal heat drift, and with $R_L$ referred to the bar width rather than a probe
spacing. It places $j_c\simeq0.1$--$0.7$~A/m, at the low end of the experimental
$0.5$--$2$~A/m of Table~\ref{tab:scales}; with the $T^{5}$ closure the same sweep shows
no instability up to $5\,\mu$A at the plateau centre. These thresholds are not
predictions. The hot branch itself is not described by the present
calculation, because a strip at tens of kelvin would smear its own density profile, which
is held frozen here. \emph{What the narrowing is and is not attributed to.} The one-sided narrowing of
Figs.~\ref{fig10}--\ref{fig13} is a class-B result: it is produced by the Hall-potential
feedback alone, with $\Te=\TL$. The thermal calculations of this section are class C and
C$'$ and do not enter it. Separating an electrostatic from a thermal contribution to
$B_c(I)$ requires the matched scans A (no feedback), B (feedback, $\Te=\TL$) and D
(feedback and temperature jointly converged); A and B are Figs.~\ref{fig3} and
\ref{fig13}, D has not been computed, and until it is, the safe statement is the one of
the abstract: finite-current electrostatic feedback produces asymmetric strips and a
one-sided narrowing within the quasilocal screening model, and a post-processed local
thermal model indicates that the narrow current-carrying strips may heat significantly.
Which loss law is closer to the truth is the question a measured $P_{\rm loss}$ inside a
strip would settle.

\section{Relation to Landauer--B\"uttiker transport}\label{sec:lb}

The channel-counting description \cite{Buttiker,Halperin} gives $G=(e^{2}/h)\sum_nT_n$
and, with backscattering suppressed, $R_{xy}=h/\nu e^{2}$. It is a limit of the present
picture: when the strips are narrow, well separated and at the boundary, each supports a few
current-carrying states, backscattering is suppressed by their separation, and counting
channels reproduces Eq.~(\ref{eq:weight}). The descriptions diverge when strips are wide,
lie in the bulk, or when the Hall potential is measured directly \cite{Ahlswede1,Ahlswede2}.
They diverge further at finite current: the widths of the two strips depend on the magnitude
and direction of $I$, so that a laterally asymmetric confinement rectifies the breakdown
\cite{Siddiki2009EPL87,Siddiki2009EPL88,Siddiki2010NJP}; and a channel of zero width does not by itself locate the heat, so that a local
electron temperature has to be supplied by an additional energy-distribution or
heat-transport description \cite{Granger2009,Altimiras2012}, whereas the strips supply
it directly. Only the quasilocal description makes the
plateau width, the position of the Hall drop, its frequency dependence, and its current
dependence into calculable quantities.

\section{Limitations and open issues}\label{sec:lim}

Disorder enters only through collision broadening; long-range fluctuations that break strips
into droplets are omitted \cite{SG2004b}. Spin splitting is carried through the gap only.
Contacts, and with them the hot spots at the current contacts \cite{Ise2005}, are idealised
away. Beyond these, four issues are specific to the present work.

\emph{The loss law.} Equation~(\ref{eq:local}) uses a single relaxation time. The
phonon-emission rate depends on $\Te$, on $B$ and on the drift velocity, and above the
phonon-emission threshold it changes character: for the transverse and longitudinal
acoustic branches of GaAs ($v_s\simeq3.0$ and $4.7$~km/s) the threshold drift velocity
corresponds to in-strip fields $E=v_sB\simeq21$--$33$~kV/m at $7$~T, which
Sec.~\ref{sec:heat} showed is reached at $I\simeq1\,\mu$A for the linear-response strip
width; beyond it the emission rate is no longer linear in $\Te-\TL$ and the constant
$\tau_0$ of Eq.~(\ref{eq:local}) is not a derived quantity, so the $0.25$~T/$\mu$A rate
of Sec.~\ref{sec:fig13} should be read as model-dependent near and above this threshold. A
microscopically derived $P_{\rm loss}(\Te,B)$ \cite{AkeraSuzuura2005,Kanamaru2006} is the
obvious refinement, and it would fix the prefactor of the $0.25$~T/$\mu$A rate of
Fig.~\ref{fig13}.

\emph{Frozen profile in the thermal solution.} The thermal fields of
Sec.~\ref{sec:valid} are computed on the equilibrium density profile with the
back-reaction of $\Te$ carried through $\sigma_l$ only; the feedback of $\Te$ on the
density itself, and the joint feedback of Hall potential and temperature that the runs of
Figs.~\ref{fig7}--\ref{fig13} include for the potential, is the calculation that would
describe the hot branch and is not done here.

\emph{Chiral heat transport.} Heat convected downstream along the strip is absent from
Eq.~(\ref{eq:balance}); its size relative to the local balance is set by
$\ell_\epsilon/L_y$ (Sec.~\ref{sec:chiral}) and it is what a two-dimensional
thermohydrodynamic extension would supply.

\emph{Thermal transport.} $\kappa_{xx}$ is tied to $\sigma_l$ by a metallic Lorenz
number, which is an assumption for activated transport in a broadened gap, and the
sensitivity to $L$ has to be checked (Sec.~\ref{sec:valid}). More seriously, in a
magnetic field the thermal and thermoelectric response is tensorial,
$\jvec=\shat\Evec-\hat\alpha\nabla T$, $\jvec_Q=T\hat\alpha\Evec-\hat\kappa\nabla T$,
and the present model keeps only $\kappa_{xx}$: the thermal Hall term $\kappa_{xy}$, the
thermoelectric coefficients $\alpha_{xx}$, $\alpha_{xy}$, and the Ettingshausen and
Peltier contributions are dropped, as is heat carried by edge or collective modes. The
edge-to-edge electron-temperature difference observed by Kawaguchi \emph{et al.}
\cite{Kawaguchi2004} is an Ettingshausen effect and cannot be reproduced by
Eq.~(\ref{eq:balance}) as written; near the plateau edge, where the strip is narrow and
$\kappa_{xx}$ is not exponentially small, the hot strip will also warm its surroundings.

\emph{The averaging kernel under current and frequency.} $\lambda=30$~nm was introduced as
a wavefunction extent at equilibrium. Whether the same kernel is legitimate for a heated
strip, or for a response function at finite $\omega$, has been assumed rather than
checked; the dynamic length $\sqrt{D/\omega}$ exceeds $\lambda$ below the terahertz range.

\emph{Finite frequency and finite amplitude together.} Sections~\ref{sec:res} and
\ref{sec:nl} are linear response at finite $\omega$ and nonlinear response at $\omega=0$.
The remaining case would require the density, and now $\Te$, to respond within the cycle. A
DC response that rectifies cannot remain linear under AC drive: if reversing the current
interchanges the strips and the hot side, a sinusoidal current must produce a
second-harmonic Hall voltage, and a thermal response with a time constant $\tau_\epsilon$
must produce a frequency-dependent onset of the plateau narrowing near
$\omega\tau_\epsilon\sim1$, in the $10^{8}$~Hz range. Both are clean tests of whether the
asymmetry and the heating share a mechanism.

\section{Conclusion}

The chain assembled here is short and each link has a stated validity condition. Canonical
action quantisation gives the Landau spectrum. Smooth confinement makes the spectrum local,
and the resulting $E_n(X)$ are Born--Oppenheimer surfaces with adiabatic parameter
$\varepsilon_{\rm ad}=\ell|\nabla V|/\hbar\omega_c$. Screening turns the local spectrum into
a self-consistent landscape of compressible regions and incompressible strips. A quasilocal
Ohm law converts that landscape into resistances, in the form
$R_H=\int\rho_Hj_y\,dx/\int j_y\,dx$ with $j_y\propto1/\rho_l$, and the same identity
places the Joule power $j_yE_y$ in the strips.

Adding a drive extends the hierarchy. At finite frequency the second adiabatic parameter
$\omega/\omega_c$ leaves the local tensor quasistatic and the correction electrostatic:
the strips keep the current, the quantisation error is second order in $\omega$, and at
kilohertz frequencies the intrinsic correction is six orders of magnitude below what AC
metrology compensates. At finite current the Hall potential feeds back on the landscape
and breaks its symmetry, and the Joule power, deposited in the strip that carries the
current and unable to leave it sideways, raises the electron temperature there. Following
Akera, that temperature is the variable that controls $\sigma_{xx}$; here its location is
supplied by the screening theory. The calculated result is a plateau that keeps its quantised value and, for the
parameters used, its high-field edge at $\nu(0)=2$, and loses width from its low-field
flank at a rate set by current --- $B_c=6.0$, $6.5$ and $7.0$~T at $1$, $3$ and
$5\,\mu$A. The electron-temperature field computed on the transport solution confirms the
first half of the heating reading: the Joule power sits in the current-carrying strip,
the strip cannot conduct it sideways, and the narrow strips of the low-field flank warm
by of order the lattice temperature at microampere currents in either closure of the
phonon loss. The second half --- whether the broad strips of the plateau centre heat, and
whether the cold strip is thermally unstable at $0.3$--$1\,\mu$A or stable to $5\,\mu$A
--- depends on the loss law inside a gapped, broadened Landau level, which this
calculation brackets but does not fix. Within that bracket the breakdown of the quantised
Hall effect is the thermal failure of the hottest strip --- a local event with a
calculable position --- and the measurement that would close the bracket is the
electron--phonon energy-loss rate of an incompressible strip.

\end{document}